\documentclass[lineno]{jfm}

\usepackage{graphicx}
\usepackage{newtxtext}
\usepackage{newtxmath}
\usepackage{natbib}
\usepackage{soul}
\usepackage{amsmath}
\usepackage{hyperref}
\usepackage[normalem]{ulem}
\def\pd{\partial}
\hypersetup{
    colorlinks = true,
    urlcolor   = blue,
    citecolor  = black,
}

\newcommand{\RomanNumeralCaps}[1]
\linenumbers

\def\pd{\partial}
\def\mch{\mathcal{H}}
\def\mcp{\mathcal{P}}

\title{Buoyancy segregation suppresses viscous fingering in horizontal displacements in a porous layer}
\author{Edward M. Hinton\aff{1}\corresp{\email{ehinton@unimelb.edu.au}} and Apoorv Jyoti\aff{2}}

\affiliation{\aff{1}School of Mathematics and Statistics, The University of Melbourne, Victoria 3010, Australia
\aff{2}Peter Cook Centre for CCS Research, School of Geography, Earth and Atmospheric Sciences, The University of Melbourne, Victoria 3010, Australia}

\begin{document}
\maketitle

\begin{abstract}
We consider the axisymmetric displacement of an ambient fluid by a second input fluid of lower density and lower viscosity in a horizontal porous layer. If the two fluids have been vertically segregated by buoyancy, the flow becomes self-similar with the input fluid preferentially flowing near the upper boundary. We show that this axisymmetric self-similar flow is stable to angular-dependent perturbations for any viscosity ratio.
The Saffman-Taylor instability is suppressed due to the buoyancy segregation of the fluids. The radial extent of the segregated current is inversely proportional to the viscosity ratio. This horizontal extension of the intrusion eliminates the discontinuity in the pressure gradient between the fluids associated with the viscosity contrast. Hence, at late times viscous fingering is shut down even for arbitrarily small density differences. The stability is confirmed through numerical integration of a coupled problem for the interface shape and the pressure gradient, and through complementary asymptotic analysis, which predicts the decay rate for each mode.
The results are extended to anisotropic and vertically heterogeneous layers. The interface may have steep shock-like regions but the flow is always stable when the fluids have been segregated by buoyancy, as in a uniform layer.
\end{abstract}

\section{Introduction}
\label{sec:intro}
Displacement flows in porous media occur in many important environmental settings including geological CO\textsubscript{2} sequestration, underground hydrogen storage, geothermal energy generation and the infiltration of sea water into fresh water aquifers. If the displacing fluid is of lesser viscosity than the ambient or displaced fluid, then the fluid-fluid interface can become unstable. This phenomenon is known as the `Saffman-Taylor instability' or as `viscous fingering' owing to the fingers of low viscosity input fluid that penetrate the ambient fluid \citep{saffman1958penetration,paterson1981radial,homsy1987viscous}. In many applications, understanding this instability is a key concern. For example, sequestered carbon dioxide is much less viscous than the host brine in the aquifer and viscous fingering may significantly reduce the fraction of the aquifer that can be accessed by the CO\textsubscript{2}, which in turn reduces the storage efficiency \citep{bachu2015review}. Multiple strategies have been developed to suppress viscous fingering including controlling the flow geometry and varying the volume flux of the input fluid \citep{al2012control,zheng2015controlling}. Significant miscibility between the fluids has also been shown to shut down fingering at later times \citep{nijjer2018dynamics,sharma2020control}.

In the case that the two fluids have different densities, as is relevant to CO\textsubscript{2} sequestration, variations in the hydrostatic pressure provide an extra horizontal force that drives (or opposes) the fluid displacement; such flows are known as `gravity currents' \citep{huppert1995gravity}. There has been much research into gravity currents in porous media, particularly in confined layers in which the ambient fluid must be displaced \citep{hesse2007gravity,juanes2010footprint,zheng2022influence}. Viscous fingering can occur at early times if the input fluid is of lower viscosity. However, it is generally assumed that at later times a stable interface between the input and ambient fluids develops, which seems to be corroborated by laboratory experiments \citep{nordbotten2006similarity,pegler2014fluid,guo2016axisymmetric}. It has also been shown that the combination of buoyancy and mixing can suppress the Saffman-Taylor instability at long times \citep{tchelepi1994viscous,riaz2003three,nijjer2022horizontal}. However, for sharp-interface gravity currents in confined porous media, stability of the gravity current solution has not yet been confirmed and the flow physics that suppresses viscous fingering is not fully understood.

\begin{figure}
\centering
\includegraphics[width=0.95\columnwidth]{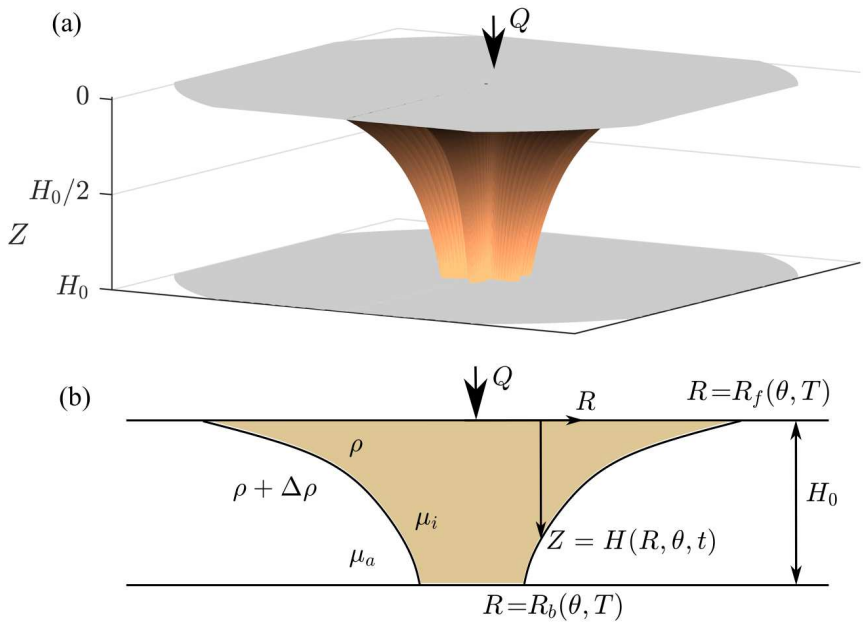}
\caption{(a) Schematic of the setup. (b) Cross-section of the setup.}
\label{fig:setupcartoon}
\end{figure}

In the present paper, we consider injection of a relatively buoyant fluid into an aquifer confined above and below by two impermeable layers. The injection source is on the upper boundary and the aquifer is initially filled with a second fluid of greater viscosity (see figure \ref{fig:setupcartoon}). We show that if the fluids have segregated owing to buoyancy, the gravity current solution is stable to both axisymmetric and angular-dependent perturbations for any values of the viscosity ratio and the input flux relative to the buoyancy velocity. 

The classical Saffman-Taylor instability is associated with the discontinuity in the pressure gradient across the fluid-fluid interface \citep{paterson1981radial}. For example, in the case of flow in a porous medium with no vertical dimension, the base radial displacement has a circular interface. The total radial flux in each fluid in a circular cross-section is given by
\begin{equation} \label{eq:classicalflux}
q_r = \frac{-2 \pi r k}{\mu}  \frac{\pd p}{\pd r},
\end{equation}
where $k$ is the permeability, $\mu$ is the viscosity, $r$ is the radial coordinate and $p$ is the pressure. This flux is continuous across the interface so the pressure gradient either side of the interface is proportional to the viscosity of the fluid there. If the pressure gradient is larger in the ambient fluid then perturbations to the interface into the ambient fluid experience a larger pressure gradient and tend to grow. Hence viscous fingers can arise if the ambient fluid is of greater viscosity. 

In a three-dimensional porous layer with fluids of different densities that have been vertically segregated by buoyancy, the input fluid will form an intrusion with large radial extent at later times. The radial flux at the interface no longer needs to be continuous because the input intrusion advances into the ambient fluid. Indeed, we will show that the vertical structure of the flow ensures that there is no destabilising pressure gradient across the interface. This stabilisation occurs for \textit{any} density difference between the two fluids that eventually leads to buoyancy segregation.

The combination of buoyancy segregation and an intrusion of large radial extent is a fundamentally different stability mechanism to the related problem of two-layer viscous gravity currents where two viscous fluids co-flow driven by gravity alone. In this case, buoyancy segregation does not guarantee stability. Instead, stability is controlled by competing contributions to the pressure gradient from buoyancy and the viscosity contrast \citep{kowal2019stability}. 
The Saffman-Taylor instability is suppressed provided that the density difference is sufficiently large. A similar phenomena occurs in unconfined flows with a longitudinal viscosity contrast \citep{kowal2021viscous}.

The stability of single-layer unconfined gravity currents (in which the aquifer is bounded only on one side by a single impermeable layer), has been investigated extensively \citep{pattle1959diffusion,grundy1982eigenvalues,newman1984lyapunov,bernoff2002linear,chertock2002stability,mathunjwa2006self}. 
 In this case, the motion of the ambient fluid is unimportant and the gravity current is not resisted by its displacement; the solutions are always stable.
The stability of confined gravity currents is different because the displacement of the more viscous ambient fluid influences the physics. For these confined flows, stability has been established for perturbations that do not depend on the transverse or azimuthal coordinate \citep{nordbotten2006similarity}. The stability analysis of both unconfined gravity currents and axisymmetric variations to confined gravity currents is significantly simplified because the driving pressure gradient in the input fluid can be eliminated from the problem. The present study of three-dimensional perturbations to confined currents requires the solution to a coupled system for both the pressure gradient and the interface shape.

The paper is structured as follows. The model for the buoyancy-segregated displacement flow is formulated in \S \ref{sec:model}. The coupled system that governs the background pressure and the interface shape is integrated numerically in \S \ref{sec:numerics} with the results demonstrating that the axisymmetric solutions are stable to angular-dependent perturbations for all values of the viscosity ratio and the density difference. In \S \ref{sec:stabilmech}, the physical mechanism that suppresses the Saffman-Taylor instability is analysed. A linear stability analysis is carried out in \S \ref{sec:linstabil} for the case of buoyancy-segregated fluids with an arbitrarily small density difference. In \S \ref{sec:vertvarperm}, we show that the stabilising effect of buoyancy segregation generalises to layers with vertically varying permeability for which the interface may contain steep shock-like regions. Concluding remarks are made in \S \ref{sec:conc}.

\section{Model} \label{sec:model}
The setup is shown in figure \ref{fig:setupcartoon}. A similar geometry was studied by \citet{nordbotten2006similarity} and \citet{guo2016axisymmetric}. Fluid of density $\rho$ and viscosity $\mu_i$ is injected into a porous medium of thickness $H_0$ and porosity $\phi$. The medium is initially filled with a second fluid of greater density $\rho + \Delta \rho$ and different viscosity $\mu_a$. The top and bottom boundaries of the layer are impermeable. The input fluid is injected with volume flux $Q$ from a point source at the upper boundary, which we take to be the origin of our coordinate system. 
The $Z$ coordinate is measured downwards from the upper boundary at which $Z=0$. We use cylindrical polar coordinates with $R$ denoting the distance from the $Z$ axis and $\theta$ denoting the angular coordinate. Time is denoted by $T$. We neglect any intermingling of the fluids and assume that there is a sharp interface at $Z=H(R,\theta,T)$ between the fluids. We also assume that the two fluids are segregated by buoyancy so that $H$ is a single-valued function.

When the input fluid has spread far from the source, the flow is predominantly in the radial direction. We may then apply the shallow flow approximation and the pressure is approximately hydrostatic,
\begin{equation}
P= \bar{P} + \rho g Z \qquad \text{for} \quad 0<Z<H,
\end{equation}
\begin{equation}
P= \bar{P} -\Delta \rho g H + (\rho+\Delta \rho) g Z \qquad \text{for} \quad H<Z<H_0,
\end{equation}
in the input and ambient fluids, respectively, where $\bar{P}(R,\theta,T)$ is the pressure at the upper boundary, $Z=0$, which is to be determined. The Darcy velocities are given by
\begin{equation}
\mathbf{U}_i = -\frac{k}{\mu_i} \nabla \bar{P},
\end{equation}
\begin{equation}
\mathbf{U}_a = -\frac{k}{\mu_a}\Big( \nabla \bar{P} - \Delta \rho g \nabla H \Big),
\end{equation}
in the input and ambient fluids, respectively.
The governing equations of the flow are
\begin{equation} \label{eq:dimgov1}
\phi \frac{\pd H}{\pd T} + \nabla \cdot \big( H \mathbf{U}_i \big) = 0
\end{equation}
and 
\begin{equation} \label{eq:dimgov2}
\nabla \cdot \big[ H \mathbf{U}_i + (H_0-H) \mathbf{U}_a  \big] = 0,
\end{equation}
which represent mass conservation integrated over the thickness of the input fluid, and mass conservation integrated over the thickness of the layer, respectively.
\par 
At $T=0$, the porous layer is entirely filled with the ambient fluid and constant-flux injection of the input fluid begins. Global mass conservation of the input fluid takes the form
\begin{equation} \label{eq:dimglobalmass}
\int_0^{2\pi} \int_0^{R_f(\theta,T)} \phi R H \, \mathrm{d} R\, \mathrm{d} \theta = QT,
\end{equation}
where $R_f(\theta,T)$ denotes the frontal contact line where $H=0$ (see figure \ref{fig:setupcartoon}b).

Eventually, the fluid-fluid interface will touch the lower boundary \citep{nordbotten2006similarity}. It is well-known that the predicted interface shape of an unconfined axisymmetric gravity current has an unphysical singularity at the origin, whilst in confined layers, the flow will always encompass the entire layer near the source, regardless of the thickness of the layer \citep{lyle2005axisymmetric,guo2016axisymmetric}. Recently, \citet{benham2022axisymmetric} carried out a detailed analysis of the flow near the source, incorporating the pressure-driven vertical flow there. Their results corroborate the conclusion that a confined layer becomes filled with input fluid at sufficiently late times \citep[see also][]{huppert2022fate}. After the fluid-fluid interface has touched the lower boundary, there is a second contact line at $R=R_b(\theta,T)$ where $H=H_0$. The flow then has three regions: there is the `fully-flooded' zone in which the input fluid fills the layer ($H=H_0$; $0\leq R < R_b(\theta,t)$); next is the partially filled zone in which the fluid-fluid interface lies between the top and bottom boundaries ($0<H<H_0$; $R_b(\theta,t)<R<R_f(\theta,t)$); finally, beyond this there is an uninvaded zone in which the ambient fluid fills the layer ($H=0$; $R>R_f(\theta,t)$); see figure \ref{fig:setupcartoon}b.

When the input fluid has filled the layer near the source, the boundary conditions are as follows.
At the source,
\begin{equation} \label{eq:dimBCsource}
H \to H_0, \qquad \frac{\pd \bar{P}}{\pd R}  \to \frac{-\mu_i Q}{2 \pi k H_0 R} \qquad \text{as} \quad R \to 0,
\end{equation}
where the latter arises from volume conservation. In the far field, the boundary condition is
\begin{equation} \label{eq:dimBCfar}
H \to 0, \qquad \frac{\pd \bar{P}}{\pd R}  \to \frac{-\mu_a Q}{2 \pi k H_0 R} \qquad \text{as} \quad R \to \infty.
\end{equation}
The dependent variables, $\bar{P}(R,\theta,T)$ and $H(R,\theta,T)$, are defined on $\theta \in [0,2 \pi)$ with periodic boundary conditions: $\bar{P}(R,0,T)=\bar{P}(R,2 \pi,T)$ and $H(R,0,T)=H(R,2 \pi,T)$. We also require that $0 \leq H \leq H_0$.
Equations \eqref{eq:dimgov1} and \eqref{eq:dimgov2} combined with global mass conservation \eqref{eq:dimglobalmass} and the boundary conditions \eqref{eq:dimBCsource}, \eqref{eq:dimBCfar} define a complete coupled system for determining the two dependent variables $\bar{P}(R,\theta,T)$ and $H(R,\theta,T)$.

Although the system derived above is sufficient to solve the problem numerically (see \S \ref{sec:numerics}), the stability investigation requires detailed treatment of the behaviour in the vicinity of the two contact lines where the interface touches the layer boundaries. Across these lines, the gradient of the interface and the gradient of the pressure at the top of the layer can be discontinuous. Indeed, it is well-known that the stability of the fluid-fluid interface for equally dense fluids is controlled by the jump in the pressure gradient at the interface (see \S \ref{sec:intro}). Therefore, we derive boundary conditions for the interface shape and the pressure gradient across each contact line.

At the trailing contact line, $R=R_b(\theta, t)$, continuity of the interface, continuity of the pressure and continuity of the flux of the input fluid take the following forms
\begin{equation} \label{eq:BCsRbfirst}
H(R=R_b^+)=H(R=R_b^-)=H_0, \qquad \big[\bar{P}\big]^+_-=0, \qquad \Bigg[ \frac{\pd \bar{P}}{\pd R} - \frac{\pd R_b}{\pd \theta} \frac{1}{R} \frac{\pd \bar{P}}{\pd \theta} \Bigg]^+_-=0,
\end{equation}
respectively, where $[\cdot]^+_-$ denotes the discontinuity in the quantity in square brackets at the contact line $R=R_b(\theta, t)$. In addition, we have a kinematic boundary condition; at the trailing contact line, the radial velocity of the ambient fluid is equal to the radial velocity of the interface, which takes the following form
\begin{equation} \label{eq:BCsRbsec}
-\frac{k}{\mu_a} \Bigg(\frac{\pd \bar{P}}{\pd R} - \Delta \rho g \frac{\pd H}{\pd R} \Bigg)\Bigg\rvert_{R=R_b^+} = \frac{\pd R_b}{\pd T} -\frac{k}{\mu_a}\frac{\pd R_b}{\pd \theta} \Bigg(\frac{1}{R}\frac{\pd \bar{P}}{\pd \theta} - \frac{\Delta \rho g}{R} \frac{\pd H}{\pd \theta} \Bigg)\Bigg\rvert_{R=R_b^+}.
\end{equation}
At the leading contact line, $R=R_f(\theta,t)$, the corresponding boundary conditions are
\begin{equation}
H(R=R_f^+)=H(R=R_f^-)=H_0, \qquad \big[\bar{P}\big]^+_-=0,
\end{equation}
\begin{equation}
\Bigg[ \frac{\pd (\bar{P}-\Delta \rho g H)}{\pd R} - \frac{\pd R_f}{\pd \theta} \frac{1}{R} \frac{\pd  (\bar{P}-\Delta \rho g H)}{\pd \theta} \Bigg]^+_-=0,
\end{equation}
\begin{equation}
-\frac{k}{\mu_i} \frac{\pd \bar{P}}{\pd R}\Bigg\rvert_{R=R_f^-} = \frac{\pd R_f}{\pd T} -\frac{k}{\mu_i R} \frac{\pd R_f}{\pd \theta} \frac{\pd \bar{P}}{\pd \theta}\Bigg\rvert_{R=R_f^-},
\end{equation}
where $[\cdot]^+_-$ denotes the discontinuity in the quantity in square brackets at the contact line $R=R_f(\theta, t)$.

\subsection{Transformed coordinate system} \label{sec:transformed}
Mass conservation of the input fluid suggests that the radial extent of the currents grows with $R^2 \sim QT/(\phi H_0)$. This motivates introducing the following transformed dimensionless coordinates
\begin{equation} \label{eq:etaandtau}
\eta = \Bigg(\frac{2 \pi \phi H_0}{Q}\Bigg)^{1/2} \frac{R}{T^{1/2}}, \qquad \tau=\log \big(T/T_{\mathrm{ref}} \big),
\end{equation}
where $T=T_{\mathrm{ref}}>0$ is a reference time, which corresponds to $\tau=0$.
We also scale the dependent variables to obtain the following dimensionless quantities
\begin{equation}
\mathcal{H}(\eta,\theta,\tau) = \frac{H(R,\theta,T)}{H_0}, \qquad \mathcal{P}(\eta,\theta,\tau)=\frac{2 \pi k H_0 \bar{P}(R,\theta,T)}{\mu_i Q}.
\end{equation}
The two mass conservation equations (\ref{eq:dimgov1}, \ref{eq:dimgov2}) become
\begin{equation} \label{eq:localmasscons_sim}
\frac{\pd \mch}{\pd \tau}-\frac{\eta}{2} \frac{\pd \mch}{\pd \eta} - \tilde{\nabla} \cdot \big( \mch \tilde{\nabla} \mcp \big) = 0,
\end{equation}
\begin{equation} \label{eq:fullmasscons_sim}
\tilde{\nabla} \cdot \Big[\big(M+(1-M)\mch\big) \tilde{\nabla} \mcp - G M (1-\mch) \tilde{\nabla} \mch   \Big] = 0,
\end{equation}
where
\begin{equation}
\tilde{\nabla}=\Bigg(\frac{\pd}{\pd (\eta \cos \theta)}, \, \frac{\pd}{\pd (\eta \sin \theta)} \Bigg)
\end{equation}
is the gradient operator with respect to the transformed coordinates and we have introduced the following two dimensionless groups
\begin{equation} \label{eq:GandM}
M=\frac{\mu_i}{\mu_a}, \qquad G= \frac{2 \pi k \Delta \rho g H_0^2}{\mu_i Q},
\end{equation}
which are the viscosity ratio and gravity number, respectively. The latter represents the magnitude of pressure gradients associated with buoyancy relative to pressure gradients associated with injection of fluid. In the case of equally dense fluids, the gravity number is $G=0$.
The boundary conditions at the source and in the far field are
\begin{equation} \label{eq:bc1_sim}
\mch \to 1, \qquad \frac{\pd \mcp}{\pd \eta}  \to \frac{-1}{\eta} \qquad \text{as} \quad \eta \to 0,
\end{equation}
and
\begin{equation} \label{eq:bc2_sim}
\mch \to 0, \qquad \frac{\pd \mcp}{\pd \eta}  \to \frac{-M^{-1}}{\eta} \qquad \text{as} \quad \eta \to \infty.
\end{equation}
The trailing contact line in the transformed coordinates is denoted by $\eta=\eta_b(\theta,\tau)$ and the boundary conditions there (\ref{eq:BCsRbfirst}, \ref{eq:BCsRbsec}) are recast as
\begin{equation} \label{eq:transBCtrail1}
\mch(\eta=\eta_b^+)=\mch(\eta=\eta_b^-)=1, \qquad \big[\mcp \big]^+_-=0, \qquad \Bigg[ \frac{\pd \mcp}{\pd \eta} - \frac{\pd \eta_b}{\pd \theta} \frac{1}{\eta} \frac{\pd \mcp}{\pd \theta} \Bigg]^+_-=0,
\end{equation}
\begin{equation}\label{eq:transBCtrail2}
-M\Bigg( \frac{\pd \mcp}{\pd \eta} - G \frac{\pd \mch}{\pd \eta} \Bigg)\Bigg\rvert_{\eta=\eta_b^+}=\frac{\pd \eta_b}{\pd \tau} +\frac{1}{2} \eta_b -M \frac{\pd \eta_b}{\pd \theta} \Bigg( \frac{1}{\eta} \frac{\pd \mcp}{\pd \theta} -\frac{G}{\eta} \frac{\pd \mch}{\pd \theta}\Bigg)\Bigg\rvert_{\eta=\eta_b^+},
\end{equation}
whilst at the leading contact line, $\eta=\eta_f(\theta,\tau)$,
\begin{equation}\label{eq:transBCfront1}
\mch(\eta=\eta_f^+)=\mch(\eta=\eta_f^-)=0, \qquad \big[\mcp \big]^+_-=0,
\end{equation}
\begin{equation}\label{eq:etafbc3}
\Bigg[ \frac{\pd (\mcp-G \mch)}{\pd \eta} - \frac{\pd \eta_b}{\pd \theta} \frac{1}{\eta} \frac{\pd (\mcp-G\mch)}{\pd \theta} \Bigg]^+_-=0,
\end{equation}
\begin{equation} \label{eq:transBCfront2}
-\frac{\pd \mcp}{\pd \eta}\Bigg\rvert_{\eta=\eta_f^-}  =\frac{\pd \eta_f}{\pd \tau} +\frac{1}{2} \eta_f -\frac{1}{\eta} \frac{\pd \eta_b}{\pd \theta}\frac{\pd \mcp}{\pd \theta}\Bigg\rvert_{\eta=\eta_f^-}.
\end{equation}
Mass conservation \eqref{eq:dimglobalmass} takes the form
\begin{equation} \label{eq:transformglmasscon}
\frac{1}{2 \pi} \int_0^{2 \pi} \int_0^{\eta_f} \eta \mch_0 \, \mathrm{d} \eta \, \mathrm{d} \theta = 1.
\end{equation}

\subsection{Self-similar axisymmetric solution}
The flow has a self-similar axisymmetric solution with \citep{nordbotten2006similarity,guo2016axisymmetric}
\begin{equation}
\mch(\eta,\theta,\tau)=\mch_0(\eta), \qquad \mcp(\eta,\theta,\tau) = \mcp_0(\eta),
\end{equation}
and
\begin{equation}
\eta_b(\theta,\tau) = \eta_{b_0}, \qquad \eta_f(\theta,\tau) = \eta_{f_0},
\end{equation}
so that the contact lines are circles, fixed in similarity space.
We are interested in analysing the stability of and convergence to these solutions in \S\S \ref{sec:numerics}, \ref{sec:stabilmech} and \ref{sec:linstabil}. In the present section, we derive a single ordinary differential equation which governs the axisymmetric self-similar interface shape and we recall an important analytical solution to this equation from \citet{nordbotten2006similarity} and \citet{guo2016axisymmetric}.
For axisymmetric flow, the pressure at the top boundary, $\mcp_0(\eta)$, can be eliminated from the problem as follows. We integrate equation \eqref{eq:fullmasscons_sim} with respect to $\eta$ and apply the source flux boundary condition \eqref{eq:bc1_sim} to obtain
\begin{equation} \label{eq:P0axi}
\frac{\mathrm{d} \mcp_0}{\mathrm{d} \eta} = \frac{-1}{\eta\big(M+(1-M)\mch_0\big)} + \frac{GM (1-\mch_0)}{\big(M+(1-M)\mch_0\big)} \frac{\mathrm{d} \mch_0}{\mathrm{d} \eta}.
\end{equation}
Upon substituting \eqref{eq:P0axi} into \eqref{eq:localmasscons_sim}, we obtain the following ordinary differential equation for $\mch_0(\eta)$,
\begin{equation} \label{eq:axisymmetricODE}
-\frac{\eta}{2} \frac{\mathrm{d} \mch_0}{\mathrm{d} \eta} + \frac{1}{\eta} \frac{\mathrm{d}}{\mathrm{d} \eta} \Bigg(\frac{\mch_0}{M+(1-M) \mch_0} \Bigg) = \frac{1}{\eta} \frac{\mathrm{d}}{\mathrm{d} \eta} \Bigg(\eta \frac{G M \mch_0(1-\mch_0)}{M+(1-M) \mch_0}\frac{\mathrm{d} \mch_0}{\mathrm{d} \eta} \Bigg).
\end{equation}
\begin{figure}
\centering
\includegraphics[width=0.95\columnwidth]{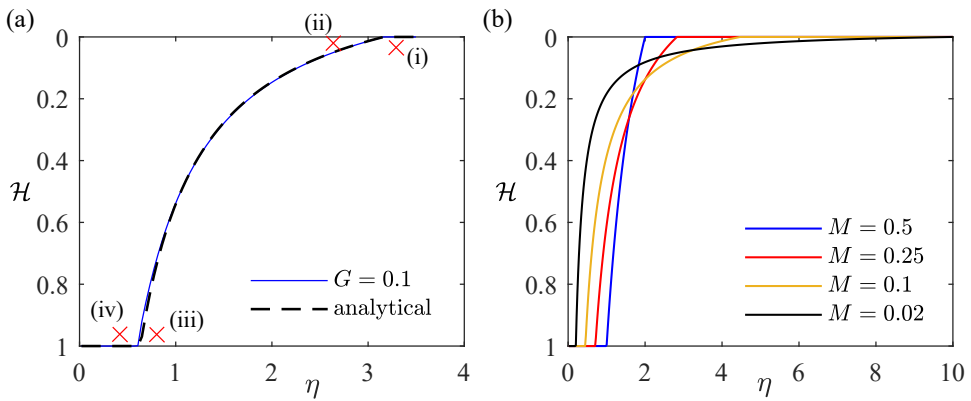}
\caption{Self-similar axisymmetric flow. (a) Interface shape for a viscosity ratio of $M=0.2$ and $G=0.1$ (blue line) and the small-$G$ approximate solution (\ref{eq:h0smallG}, black dashed line).  (b) Small-$G$ approximate interface shapes \eqref{eq:h0smallG} for four values of $M$.}
\label{fig:viscshape}
\end{figure}
This axisymmetric system can be solved numerically for $\mch_0(\eta)$ as described in \citet{guo2016axisymmetric} for a wide range of $M$ and $G$. For $G\gg 1$ (buoyancy-dominated flow), the input fluid forms a thin current near the top of the layer and the displacement of the ambient fluid is unimportant. The flow is effectively `unconfined' and single phase. Hence, we do not expect Saffman-Taylor instabilities in this regime. Instead, we focus on the case of relatively larger input flux ($G \leq \mathcal{O}(1)$) and a less viscous input fluid ($M<1$) for which stability is not well-understood. For $G \ll 1$ and $M<1$, \eqref{eq:axisymmetricODE} has an approximate analytic solution. We neglect the right-hand side of \eqref{eq:axisymmetricODE}. Although this removes the highest derivative of $\mch_0$, the boundary conditions \eqref{eq:transBCfront2} and \eqref{eq:transBCtrail2} become independent of $\mch_0$ for $G\ll 1$ and simply give the locations of the contact points so the problem remains complete.
Integrating the left-hand side of \eqref{eq:axisymmetricODE} furnishes the following interface shape \citep{guo2016axisymmetric}
\begin{align}
\mch_0&=1,  &   0<&\eta< \eta_{b_0},\\
\mch_0 &= \frac{(2M)^{1/2}}{(1-M)\eta} -\frac{M}{1-M}, &  \eta_{b_0}<&\eta< \eta_{f_0}, \label{eq:h0smallG} \\
\mch_0&=0, &  \eta_{f_0}<&\eta,
\end{align}
where the contact lines are given by 
\begin{equation}
\eta_{b_0}=\sqrt{2M} \qquad and \qquad \eta_{f_0} =\sqrt{2/M}.
\end{equation}
Note that these provide inner bounds on the locations of the contact lines for general $G> 0$ provided that $M<1$:
\begin{equation} \label{eq:contactpointineq}
\eta_{b_0} \leq \sqrt{2M}, \qquad \eta_{f_0} \geq \sqrt{2/M},
\end{equation}
with equality as $G \to 0^+$. These inequalities arise because buoyancy acts to extend the interface (its effect is proportional to $G$). 

The corresponding pressure at the top of the layer takes the form
\begin{align}
\mcp_0&=-1-\log\Bigg(\frac{\eta}{(2M)^{1/2}}\Bigg)  &   0<&\eta< \eta_{b_0},\\
\mcp_0 &= -\frac{\eta}{(2M)^{1/2}}, &  \eta_{b_0}<&\eta< \eta_{f_0}, \\
\mcp_0&=-M^{-1} -M^{-1} \log\Bigg(\frac{\eta}{(2/M)^{1/2}}\Bigg) &  \eta_{f_0}<&\eta,
\end{align}
where $\mcp_0(\eta)$ is defined up to addition of an arbitrary constant.
\par
The numerical solution to \eqref{eq:axisymmetricODE} for $\mch_0$ is plotted as a blue line in figure \ref{fig:viscshape}a for $G=0.1$ and $M=0.2$. The small-$G$ analytical solution \eqref{eq:h0smallG} is shown as a black dashed line and there is good agreement with the numerical result. Figure \ref{fig:viscshape}b shows the analytical solutions for various values of $M<1$. When the input fluid is of relatively lower viscosity (smaller $M$), it forms a finger that intrudes further into the ambient fluid at the top of the layer. 

The axisymmetric analytic solution \eqref{eq:h0smallG} is relevant to the regime in which pressure gradients associated with buoyancy are weak compared to those associated with injection ($G \ll 1$). It is worth considering the behaviour in the limit $G \to 0$, corresponding to equally dense fluids and no buoyancy force. As $G \to 0$, the axisymmetric analytic solution \eqref{eq:h0smallG} becomes a more accurate approximation of the numerical solution, with the input fluid preferentially flowing near the upper boundary (figure \ref{fig:viscshape}). However, $G=0$ corresponds to no buoyancy segregation and so the input fluid should have no preference for the top of the layer. Indeed, in the case of equally dense fluids or an arbitrarily thin layer ($G=0$), and $M<1$, one would observe classical Saffman-Taylor fingering with angular dependence and no vertical preference \citep{paterson1981radial}. The derivation of the axisymmetric analytical solution \eqref{eq:h0smallG} implicitly assumes that the input and ambient fluid have been segregated by buoyancy, which requires $G>0$ and takes a time proportional to $G^{-1}$. Hence, the case of $G=0$ has qualitatively different late-time behaviour to the case of small but non-zero $G$ and the limit $G \to 0$ is singular. In this paper, we analyse the stability when $G$ is small but positive; we exclude the case $G=0$ for which buoyancy has no effect.
\par
For $G$ small but non-zero, the interfaces in figure \ref{fig:viscshape}b occur provided that (i) buoyancy segregation has occurred and (ii) the radial extent of the flow is much greater than the layer thickness so that the shallow approximation applies. For $M<1$, in dimensional terms, the former requires
\begin{equation} \label{eq:segregatetime}
T \gg \frac{\mu_a H_0}{\Delta \rho g k},
\end{equation}
whilst the latter requires
\begin{equation} \label{eq:shallowtime}
T \gg \frac{2 \pi \phi H_0^3}{Q}.
\end{equation}
Prior to buoyancy segregation, Saffman-Taylor fingering can occur. The present paper is concerned with the post-segregation stability of the axisymmetric solutions to \eqref{eq:axisymmetricODE} to angular-dependent perturbations with $\Delta \rho>0$.
For $G\ll 1$, the first condition \eqref{eq:segregatetime} implies the second \eqref{eq:shallowtime}. Notice that for small density differences, $\Delta \rho$, the time for buoyancy segregation is very large. It is also important to note that in an anistropic porous medium, the permeability $k$ that appears in segregation timescale \eqref{eq:segregatetime} will depend only on the vertical permeability, whereas the gravity number (and hence the self-similar solutions) depends only on horizontal permeability \citep[cf.][]{benham2022axisymmetric}. The results we derive in this paper concerning flow stability extend to porous layers that have different vertical and horizontal permeabilities (as is common in many aquifers); see also \S \ref{sec:vertvarperm}. The case of more complicated anisotropy such as different horizontal permeabilities in different directions is beyond the scope of the present work.
\par
For simplicity, we have modelled the case where the source of input fluid is located at a point on the upper boundary. However, the analysis and results of this paper apply for any location of the injection source (for example, at the lower boundary, or over a vertical line within the layer). This is because the input fluid will always eventually fill the thickness of the layer near the source. Subsequently, the flow becomes predominantly radial.  There will, of course, be different early-time transient behaviour for different source locations but once \eqref{eq:segregatetime} and \eqref{eq:shallowtime} apply, the exact source location becomes unimportant.

\section{Numerical results} \label{sec:numerics}
\begin{figure}
\centering
\includegraphics[width=0.95\columnwidth]{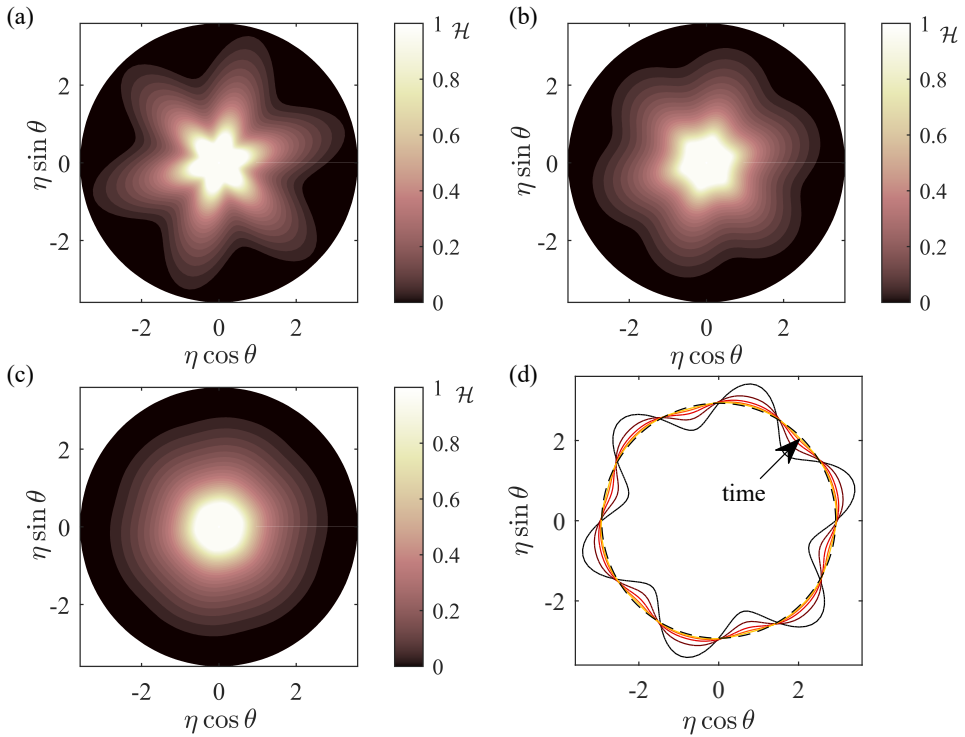}
\caption{Suppression of mode six fingers for $G=0.1$ and $M=0.2$. (a) Initial condition at the reference time ($\tau=0$) for the flow thickness, $\mch$, in transformed coordinates. (b,c) The flow thickness at $\tau=1.25$ and $\tau=4$.  (d) The evolution of the contact line, where $\mch=0$.}
\label{fig:fingersuppression}
\end{figure}

\begin{figure}
\centering
\includegraphics[width=0.95\columnwidth]{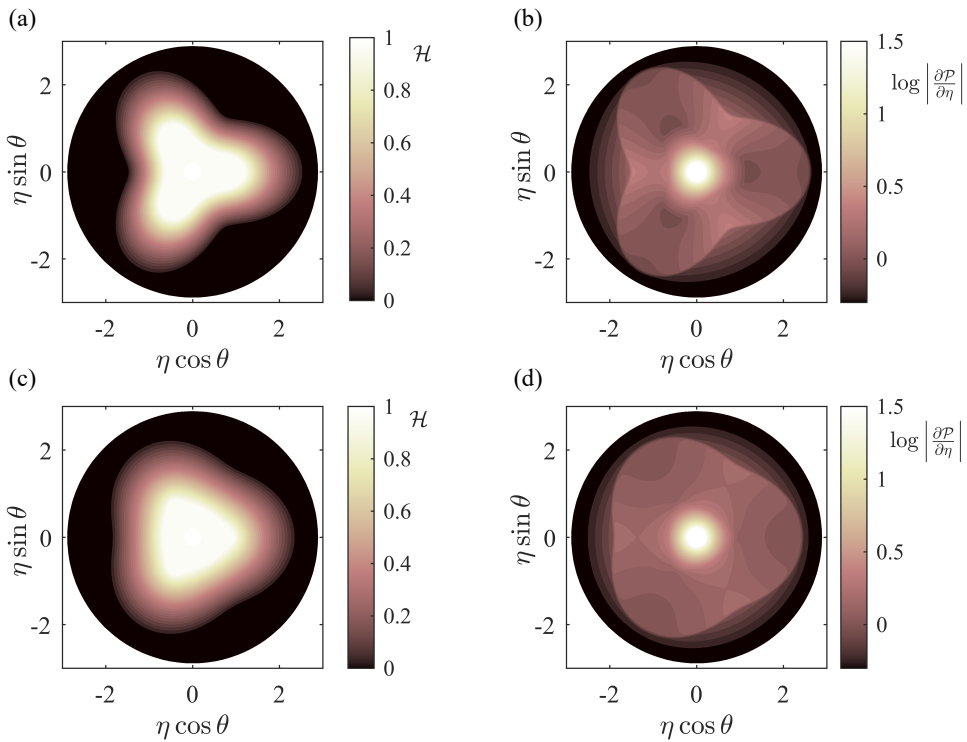}
\caption{Suppression of mode three fingers for $G=0.5$ and $M=0.5$. (a) Initial condition ($\tau=0$) for the interface shape. (b) Initial condition ($\tau=0$) for the pressue gradient (log of its magnitude is shown). (c,d) Corresponding plots at $\tau=1$. Radial cross-sections of these results are shown in figure \ref{fig:radialsupressn3}.}
\label{fig:fingersupressn3}
\end{figure}

\begin{figure}
\centering
\includegraphics[width=0.95\columnwidth]{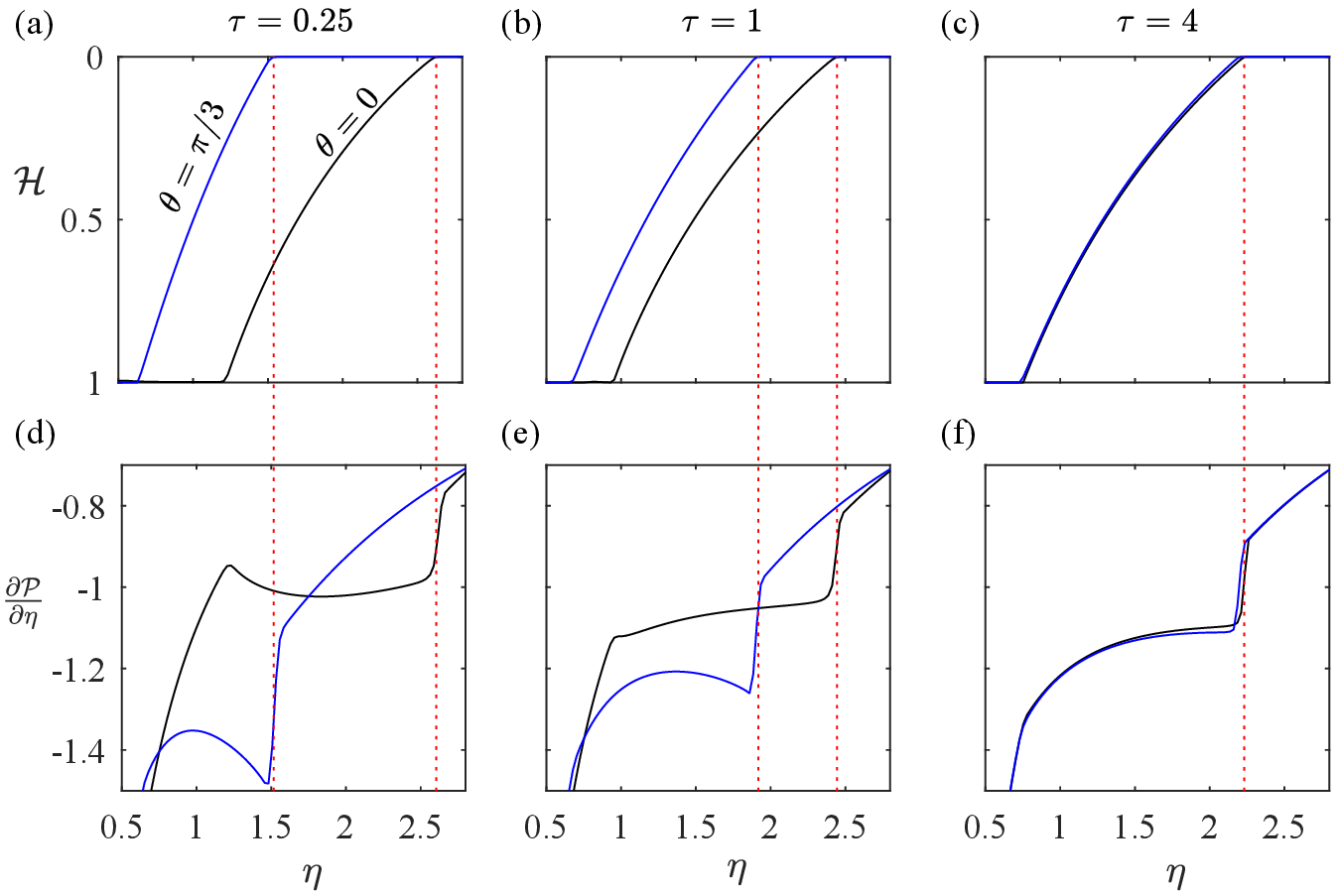}
\caption{Suppression of mode three fingers for $G=0.5$ and $M=0.5$ (same parameters and initial condition as figure \ref{fig:fingersupressn3}). Radial cross-sections at $\theta=0$ (black lines) and $\theta=\pi/3$ (blue lines) of the flow thickness, $\mathcal{H}$, and radial pressure gradient at the upper boundary $\partial \mathcal{P}/\partial \eta$, at three times: (a,d) $\tau=0.25$, (b,e) $\tau=1$, (c,f) $\tau=4$. The red dotted lines denote the locations of the frontal contact line, $\eta_f$, at $\theta=0$ and $\theta=\pi/3$.}
\label{fig:radialsupressn3}
\end{figure}

In order to investigate the response of the flow to perturbations with angular dependence, we develop a numerical method to solve the coupled system for the evolution of the fluid-fluid interface, $\mch(\eta, \theta, \tau)$, and the pressure at the upper boundary, $\mcp(\eta, \theta, \tau)$. This consists of equations \eqref{eq:localmasscons_sim} and \eqref{eq:fullmasscons_sim} with boundary conditions \eqref{eq:bc1_sim} and \eqref{eq:bc2_sim}, and an appropriate initial condition. We solve this system numerically using a finite-difference scheme, details of which are given in Appendix \ref{app:numerics}.

The numerical method is applied to a wide range of axisymmetric and non-axisymmetric initial conditions over many values of the parameters, $G$ and $M$. In all cases, the solution converges to the axisymmetric similarity solution. This demonstrates both the veracity of the numerical method and that the similarity solution provides the intermediate asymptotics for many initial conditions \citep[cf.][]{mathunjwa2006self}. Examples are shown in figures \ref{fig:fingersuppression} and \ref{fig:fingersupressn3} illustrating how mode six and mode three fingers are suppressed even though the input fluid is of lesser viscosity than the ambient. We find that any perturbation is suppressed, including radial perturbations, provided that the fluids remain buoyancy segregated (if the interface becomes nonmonotonic as a function of vertical coordinate then the model breaks down).

The numerical results also demonstrate that there is a jump in the radial pressure gradient at the upper boundary across the leading contact line (see figure \ref{fig:radialsupressn3}). This pressure gradient jump is associated with the density difference between the two fluids and the discontinuity vanishes as $G \to 0$. The discontinuity has a stabilising effect because the magnitude of the pressure gradient is smaller in the ambient fluid (see \S \ref{sec:stabilmech}). However, this discontinuity in the hydrostatic pressure is not required for the interface to be stable. Indeed, in \S \ref{sec:linstabil} we show that the interface is stable even in the limit $G \to 0$.

\section{Stability mechanism} \label{sec:stabilmech}
In the present section, we describe the mechanism by which buoyancy segregation suppresses viscous fingering. In the classical case of radial flow with no vertical dimension (discussed in \S \ref{sec:intro}), the pressure gradient either side of the fluid-fluid interface is proportional to the fluid viscosity owing to continuity of the radial flux, which drives the well-known instability. 

We now analyse the more complicated three-dimensional case in which the fluids have different densities and are segregated by buoyancy. Between the contact lines, the pressure gradient in the input fluid is $\mathrm{d} \mcp_0/\mathrm{d} \eta$ and the pressure gradient in the ambient fluid is $\mathrm{d} (\mcp_0-G\mch_0)/\mathrm{d} \eta$, where the $-G \mch_0$ contribution arises from the density difference between the fluids. Since both $\mathrm{d} \mcp_0/\mathrm{d} \eta$ and $\mathrm{d} \mch_0/\mathrm{d} \eta$ are negative, the magnitude of the pressure gradient is larger in the input fluid (in contrast to the classical case with no vertical dimension), which implies that between the contact lines the interface is stable for $G>0$. The stability at the contact lines must be treated separately because there is a discontinuity in $\mathrm{d} \mcp_0/\mathrm{d} \eta$ and $\mathrm{d} \mch_0/\mathrm{d} \eta$ between the input and ambient fluids as shown in figure \ref{fig:radialsupressn3}. We consider the driving pressure gradients in the input and ambient fluids either side of the contact lines (these locations are labelled (i)-(iv) in figure \ref{fig:viscshape}a).

At location (i), just ahead of the leading contact line, at $\eta=\eta_{f_0}^+$, the interface gradient and pressure gradient at the top boundary are
\begin{equation} \label{eq:loc1}
 \frac{\mathrm{d} \mch_0}{\mathrm{d} \eta} =0, \qquad \frac{\mathrm{d} \mcp_0}{\mathrm{d} \eta} = -\frac{1}{M \eta_{f_0}}.
\end{equation}
The latter is calculated from \eqref{eq:P0axi}.
The pressure gradient driving the ambient fluid at (i) is 
\begin{equation} \label{eq:dpdetaambienti}
\frac{\mathrm{d} (\mcp_0-G \mch_0)}{\mathrm{d} \eta}=-\frac{1}{M \eta_{f_0}}.
\end{equation}
At location (ii), just behind the leading contact line, $\eta=\eta_{f_0}^-$, the interface and pressure gradients are
\begin{equation} \label{eq:dpdetaii}
\frac{\mathrm{d} \mch_0}{\mathrm{d} \eta} =\frac{1}{\eta_{f_0} M G} -\frac{\eta_{f_0}}{2G}, \qquad \frac{\mathrm{d} \mcp_0}{\mathrm{d} \eta} = -\frac{\eta_{f_0}}{2},
\end{equation}
The former is calculated by taking $\mch \to 0$ in \eqref{eq:axisymmetricODE}.
There is a discontinuity in the pressure gradient at the top boundary, $\mathrm{d} \mcp_0/\mathrm{d} \eta$ across $\eta=\eta_{f_0}$, which can be observed in the numerical results in figure \ref{fig:radialsupressn3}. This discontinuity vanishes as $G \to 0$ even for fluids with different viscosities; the jump in the pressure gradient associated with the viscosity contrast is eliminated by the radial extension of the interface.
The pressure gradient driving the input fluid at (ii) is given by \eqref{eq:dpdetaii}b, and its magnitude is greater than (or equal to) the magnitude of the pressure gradient in the ambient fluid \eqref{eq:dpdetaambienti} because $\eta_{f_0} \geq \sqrt{2/M}$ (see \ref{eq:contactpointineq}). Hence small perturbations to the contact line decay because they experience unfavourable pressure gradients, which suggests interfacial stability provided that $G>0$. The driving pressure gradients either side of the contact line are equal in the limit $G \to 0$; this case is discussed in more detail in \S \ref{sec:linstabil}.

Similar analysis applies at the trailing contact line.
At location (iii), just downstream of the trailing contact line, $\eta=\eta_{b_0}^+$, the interface gradient, the pressure gradient at the upper boundary, and the pressure gradient driving the ambient fluid are given by
\begin{equation}
\frac{\mathrm{d} \mch_0}{\mathrm{d} \eta} =\frac{\eta_{b_0}}{2 M G} - \frac{1}{G \eta_{b_0}}, \qquad \frac{\mathrm{d} \mcp_0}{\mathrm{d} \eta} = -\frac{1}{\eta_{b_0}}, \qquad \frac{\mathrm{d} (\mcp_0-G \mch_0)}{\mathrm{d} \eta}=- \frac{\eta_{b_0}}{2M},
\end{equation}
respectively.
At location (iv), just behind the trailing contact line, $\eta=\eta_{b_0}^-$, the interface gradient and the pressure gradient at the upper boundary (which drives the input fluid) are given by
\begin{equation} \label{eq:loc4}
	\frac{\mathrm{d} \mch_0}{\mathrm{d} \eta} =0, \qquad \frac{\mathrm{d} \mcp_0}{\mathrm{d} \eta} = -\frac{1}{\eta_{b_0}}.
\end{equation}
The magnitude of the pressure gradient driving the input fluid, $1/\eta_{b_0}$, is greater than (or equal to) the magnitude of the pressure gradient driving the ambient fluid, $\eta_{b_0}/(2M)$ since $\eta_{b_0} \leq \sqrt{2M}$ with equality when $G=0$. This suggests that the trailing contact line is stable to small perturbations for $G>0$. 

It is worth noting that for small $G$, at the location where $\mch_0=0$, the pressure gradient in the ambient fluid \eqref{eq:loc1}b is proportional to the relative viscosity of the ambient fluid divided by the radial distance and similarly at $\mch_0=1$, the pressure gradient in the input fluid \eqref{eq:loc4}b is proportional to the relative viscosity of the input fluid divided by the radial distance. This dependence on the relative viscosity and radial location is as in the classical case (see \ref{eq:classicalflux}). However, the key difference is that, in contrast to the classical instability, the contact lines are separated so that the radial coordinate is different at $\mch_0=0$ and $\mch_0=1$. The change in the pressure gradient between location (iv) and location (i) (figure \ref{fig:viscshape}a) is controlled by the viscosity contrast (relatively larger ambient viscosity leads to a larger increase in the magnitude of the pressure gradient) and how far apart the contact lines are, which reduces the magnitude of the pressure gradient downstream. These two effects act in opposition. If the intrusion of input fluid has a relatively short radial extent then the flow is unstable and the intrusion grows radially until the distance between the contact lines acts to stabilise the interface. Hence the self-similar axisymmetric solutions have larger radial extent at smaller viscosity ratios.  In summary, the radially extensive intrusion of the interface along the upper boundary associated with buoyancy segregation counteracts the destabilising discontinuity in the pressure gradient associated with the viscosity contrast. The vertically-segregated radial intrusion could be thought of as a single axisymmetric viscous finger, which is stable to angular-dependent fingers.

We have shown that it is not gravity that acts against the pressure gradient to stabilise the contact lines. Indeed, the radial extent of the interface is insensitive to the size of $G$ for small $G$. The role of gravity is simply to segregate the fluids and stability occurs after the segregation has occurred.
 In section \ref{sec:linstabil}, we show that provided that buoyancy has segregated the fluids, the small-$G$ axisymmetric self-similar solutions are linearly stable. This formally confirms the results in the present section.

\section{Linear stability for small density difference ($G \ll 1$)} \label{sec:linstabil}
In the present section, we demonstrate the linear stability of the $G \ll 1$ axisymmetric self-similar solutions (given by \ref{eq:h0smallG}). Although buoyancy does not appear in the form of the solution, some buoyancy is required to segregate the fluids. Confirming the stability of the $G\ll 1$ solution will demonstrate stability for $G > 0$ as discussed in \S \ref{sec:stabilmech}. The linear stability analysis also gives the rate of decay of each mode and its dependence on the viscosity ratio.

We consider $\theta$-dependent perturbations to the axisymmetric self-similar solutions of the form
\begin{align}
\mch&=\mch_0(\eta) + \epsilon \mch_1(\eta) e^{\sigma \tau + i n \theta}, \\
\eta_f &=\eta_{f_0} + \epsilon \eta_{f_1} e^{\sigma \tau + i n \theta}, \\
\eta_b &=\eta_{b_0} + \epsilon \eta_{b_1} e^{\sigma \tau + i n \theta}, \\
\mcp &= \mcp_0(\eta) +\epsilon \mcp_1(\eta) e^{\sigma \tau + i n \theta},
\end{align}
where $\epsilon \ll 1$ and $n$ is an integer. We seek to determine the stability of these perturbations. Note that the perturbation corresponds to $e^{\sigma \tau} \sim T^{\sigma}$ in terms of real time $T$ (see \ref{eq:etaandtau}).
\par
Often linear stability analyses of viscous gravity currents requires rescaling the spatial domain owing to the singular behaviour of the interface at the contact lines \citep[e.g.][]{mathunjwa2006self,kowal2019stability}. For the present porous gravity current, the interface is linear at the contact lines so such a transformation is not required. Instead, we linearise the boundary conditions about the leading order locations of the contact lines, $\eta_f=\eta_{f_0}$ and $\eta_b=\eta_{b_0}$. We consider perturbations with $\theta$-dependence ($n \geq 1$) first as the stability in this case has not yet been investigated. Note that global mass conservation of the input fluid \eqref{eq:transformglmasscon} is identically satisfied by the form of the perturbations for $n \geq 1$. The case of axisymmetric perturbations ($n=0$) is included at the end of this section for completeness.

In the single-phase regions (upstream of the trailing contact line and downstream of the leading contact line), the pressure satisfies Laplace's equation and the hence pressure perturbation is given by the solution to
\begin{equation}
\frac{1}{\eta} \frac{\mathrm{d}}{\mathrm{d} \eta}\Bigg( \eta \frac{\mathrm{d} \mcp_1}{\mathrm{d} \eta} \Bigg) - \frac{n^2}{\eta^2} \mcp_1 = 0.
\end{equation}
Given that the pressure perturbation remains finite as $\eta \to \infty$ and as $\eta \to 0$, the solution in the single-phase regions is
\begin{equation} \label{eq:p1ups}
\mcp_1=c_n \eta^n \qquad \text{for} \qquad \eta< \eta_{b_0}
\end{equation}
\begin{equation} \label{eq:p1down}
\mcp_1=d_n \eta^{-n} \qquad \text{for} \qquad \eta > \eta_{f_0}
\end{equation}
where $c_n$ and $d_n$ are constants. 

After some algebra, the linearised governing equations \eqref{eq:localmasscons_sim}, \eqref{eq:fullmasscons_sim} in the interface region ($\eta_{b_0}<\eta<\eta_{f_0}$) become
\begin{equation}
(M-1) \frac{\mathrm{d}}{\mathrm{d} \eta} \big(\eta \mch_1 \big) + 2M \Bigg( \frac{\mathrm{d}^2 \mcp_1}{\mathrm{d} \eta^2} - \frac{n^2}{\eta^2} \mcp_1 \Bigg) =0,
\end{equation}
\begin{equation} \label{eq:h1p1rel}
(1-M)(\sigma +1/2) \mch_1 = -\frac{M}{\eta} \frac{\mathrm{d} \mcp_1}{\mathrm{d} \eta}.
\end{equation}
We can then eliminate $\mch_1$ and obtain a single equation for $\mcp_1$ in the interface region between the contact lines:
\begin{equation} \label{eq:p1govsimp}
\Bigg(\frac{\sigma +1}{\sigma +1/2}\Bigg) \frac{\mathrm{d}^2 \mcp_1}{\mathrm{d} \eta^2} - \frac{n^2}{\eta^2} \mcp_1 = 0.
\end{equation}
\begin{figure}
\centering
\includegraphics{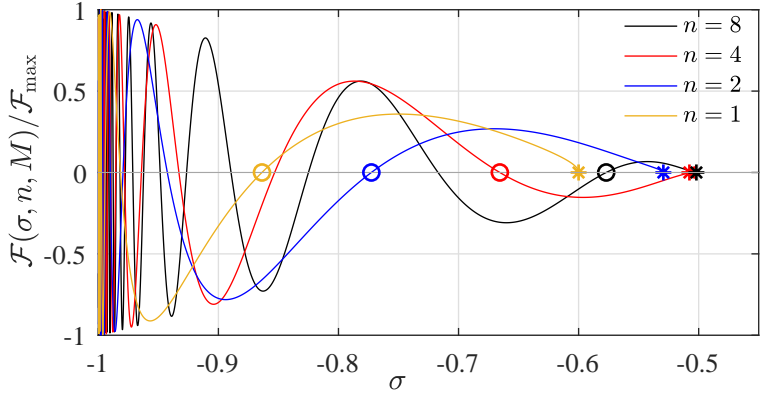}
\caption{The dispersion function, $\mathcal{F}(\sigma,n,M)$ \eqref{eq:dispersionfundef} plotted versus the growth rate $\sigma$ for $M=0.5$. It has been normalised by its maximum value over the range of $\sigma$, $\mathcal{F}_{\mathrm{max}}$. The zeros of the curves correspond to admissible decay rates $\sigma$ \eqref{eq:dispersion1} with the exception of the largest zero (shown as stars). The next largest zero (shown as circles) will be the decay rate observed (further details in the text).}
\label{fig:dispersiondifference}
\end{figure}
To determine the pressure perturbation in the interface region, we solve \eqref{eq:p1govsimp} with three boundary conditions at each contact line, which arise from the linearised version of continuity of the pressure, continuity of the flux, and the kinematic boundary condition (see \ref{eq:transBCtrail1}, \ref{eq:transBCtrail2}, \ref{eq:transBCfront1}, \ref{eq:transBCfront2}). At $\eta=\eta_{b_0}^+$, these boundary conditions take the form
\begin{align}
\mcp_1\rvert_{\eta_{b_0}^+} &= c_n \eta_{b_0}^n, \\
\frac{\mathrm{d} \mcp_1}{\mathrm{d} \eta} \bigg\rvert_{\eta_{b_0}^+} &= \frac{\eta_{b_1}}{\eta_{b_0}^2} + n c_n \eta_{b_0}^{n-1},\\
(\sigma+1/2) \eta_{b_1} &= - M\frac{\mathrm{d} \mcp_1}{\mathrm{d} \eta} \bigg\rvert_{\eta_{b_0}^+} =-M \Bigg(\frac{\eta_{b_1}}{\eta_{b_0}^2} + n c_n \eta_{b_0}^{n-1}\Bigg), \label{eq:kinlintrail}
\end{align}
where we have used the upstream behaviour \eqref{eq:p1ups}.
At $\eta=\eta_{f_0}^-$, the analogous boundary conditions take the form
\begin{align}
\mcp_1\rvert_{\eta_{f_0}^-} &= d_n \eta_{f_0}^{-n}, \\
\frac{\mathrm{d} \mcp_1}{\mathrm{d} \eta} \bigg\rvert_{\eta_{f_0}^-}&=\frac{\eta_{f_1}}{2} - n d_n \eta_{f_0}^{-n-1}, \\
(\sigma+1/2) \eta_{f_1} &= -\frac{\mathrm{d} \mcp_1}{\mathrm{d} \eta} \bigg\rvert_{\eta_{f_0}^-}=-\Bigg(\frac{\eta_{f_1}}{2} - n d_n \eta_{f_0}^{-n-1}\Bigg), \label{eq:kinlinlead}
\end{align}
where we have used the downstream behaviour \eqref{eq:p1down}.
The system for $\mcp_1$ between the contact lines is linear and has six boundary conditions with six unknown constants: $c_n$, $d_n$, $\eta_{b_1}$, $\eta_{f_1}$ and two constants arising from solving the ordinary differential equation \eqref{eq:p1govsimp}. The system is an eigenvalue problem for the growth rate $\sigma$. The general solution to equation  \eqref{eq:p1govsimp} is characterised by the value of $\sigma$ relative to the critical value
\begin{equation}
\sigma_c = \frac{-2n^2 -1}{4n^2 +1}.
\end{equation}
\begin{figure}
\centering
\includegraphics[width=0.95\columnwidth]{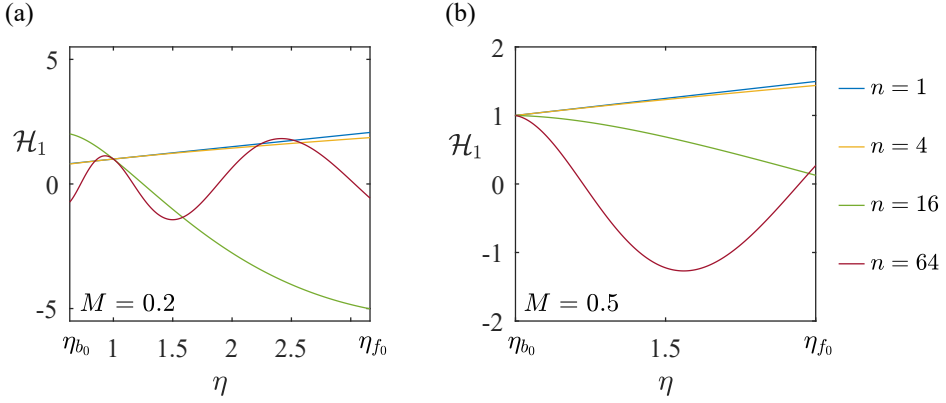}
\caption{Eigenfunctions for the interface perturbation, $\mch_1$, where we have set $\mch_1(\eta=1)=1$. (a) $M=0.2$ and (b) $M=0.5$. The eigenfunctions are calculated from \eqref{eq:h1p1rel} and \eqref{eq:eigefuns}. The $x$ axis is $\eta_{b_0}< \eta < \eta_{f_0}$, where in (a) $\eta_{b_0}=0.63$, $\eta_{f_0}=3.16$ and in (b) $\eta_{b_0}=1$, $\eta_{f_0}=2$.}
\label{fig:eigen}
\end{figure}
This critical value of $\sigma$ corresponds to a repeated root in the auxiliary equation for the ordinary differential equation \eqref{eq:p1govsimp}. There are no non-trivial solutions that satisfy all the boundary conditions for $\sigma\geq \sigma_c$. For $-1<\sigma<\sigma_c$, the solution takes the form
\begin{equation} \label{eq:eigefuns}
\mcp_1 = \eta^{1/2} \big[ a_1 \cos( \alpha \log \eta) + a_2 \sin( \alpha \log \eta)\big], 
\end{equation}
where
\begin{equation}
\alpha= \sqrt{-\frac{1}{4}-\bigg(\frac{\sigma+1/2}{\sigma +1} \bigg) n^2},
\end{equation}
and $a_1$ and $a_2$ are constants. We apply the six boundary conditions to obtain the following dispersion relation for $\sigma$
\begin{equation} \label{eq:dispersion1}
\mathcal{F}(\sigma,n,M) =0,
\end{equation}
where
\begin{equation} \label{eq:dispersionfundef}
\begin{aligned} 
\mathcal{F}&(\sigma,n,M)= \\
\Big[&\big(\sigma+ \textstyle{\frac{1}{2}}\big)n s_b -(\sigma+1)\big(\textstyle{\frac{1}{2}}s_b+\alpha c_b\big)\Big] \Big[ \big(\sigma+\textstyle{\frac{1}{2}}\big)n c_f+(\sigma+1)\big(\textstyle{\frac{1}{2}}c_f-\alpha s_f\big)\Big]-\\
\Big[&\big(\sigma+\textstyle{\frac{1}{2}} \big)n s_f +(\sigma+1)\big(\textstyle{\frac{1}{2}}s_f+\alpha c_f\big)\Big] \Big[ \big(\sigma+\textstyle{\frac{1}{2}}\big)n c_b-(\sigma+1)\big(\textstyle{\frac{1}{2}}c_b-\alpha s_b\big)\Big], 
\end{aligned}
\end{equation}
and
\begin{align}
s_b=&\sin(\alpha \log \eta_{b_0}), & s_f=&\sin(\alpha \log \eta_{f_0}), \\
c_b=&\cos(\alpha \log \eta_{b_0}), & c_f=&\cos(\alpha \log \eta_{f_0}).
\end{align}
The locations of the contact lines, $\eta_{b_0}$ and $\eta_{f_0}$, are functions of the viscosity ratio $M$ and so the dispersion equation \eqref{eq:dispersion1} also depends on $M$.
\par
The dispersion equation \eqref{eq:dispersion1} can be solved to obtain the growth rate, $\sigma$ for any $n \geq 1$ and $0<M<1$. The corresponding eigenfunctions are given by \eqref{eq:eigefuns}. Since $\sigma < \sigma_c<0$, the growth rate is always negative and hence the axisymmetric similarity solutions are linearly stable. 

The dispersion equation \eqref{eq:dispersion1} can have multiple solutions. The function $\mathcal{F}(\sigma,n,M)$ is plotted against $\sigma$ in figure \ref{fig:dispersiondifference} for $M=0.5$ and four values of $n$. For each $n$, the largest zero for $\sigma$ is $\sigma=\sigma_c$ (these solutions are indicated by stars in figure \ref{fig:dispersiondifference}). However, the case $\sigma=\sigma_c$ corresponds to a trivial solution for $\mcp_1$, which is not admissible. The next largest solution $\sigma$ of \eqref{eq:dispersion1} corresponds to the slowest decaying non-trivial solution and these values are indicated by circles in figure \ref{fig:dispersiondifference}. We expect this to be the decay rate that is observed and we take this solution as the correct value of $\sigma$.

The eigenfunctions \eqref{eq:eigefuns} that correspond to this correct solution $\sigma$ are defined up to a multiplicative constant. The form of the interfacial perturbation, $\mch_1$ for each eigenfunction \eqref{eq:eigefuns} can be calculated from \eqref{eq:h1p1rel}. These interfacial perturbation eigenfunctions are shown in figure \ref{fig:eigen} for different values of the viscosity ratio, $M$ and the mode, $n$. The shapes are more oscillatory at higher values of $n$.

\begin{figure}
\centering
\includegraphics{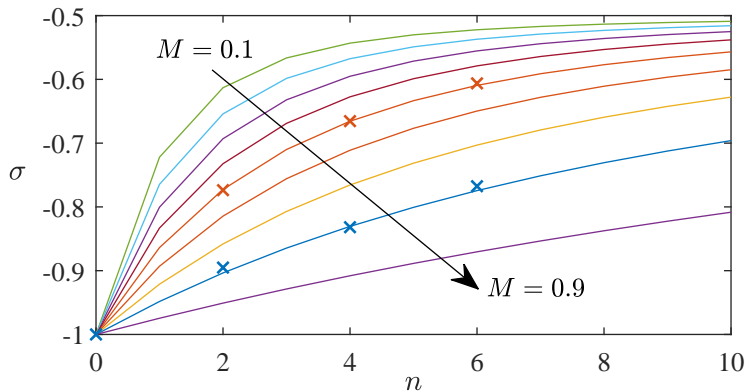}
\caption{The decay rate, $\sigma$, for $M=0.1, 0.2, 0.3, 0.4, 0.5, 0.6, 0.7,0.8, 0.9$ and $0 \leq n \leq 10$. The crosses show the decay rate observed from our numerical results (see Appendix \ref{app:numericsdecay}).}
\label{fig:decayrates}
\end{figure}

The growth rate, $\sigma$, calculated for different values of $M$ and $n$ using the dispersion relation \eqref{eq:dispersion1} is plotted using continuous lines in figure \ref{fig:decayrates}. The rate of decay is slower for larger values of $n$ and smaller viscosity ratios, $M$. 
The crosses in figure \ref{fig:decayrates} denote the decay rate calculated from the numerical method for $M=0.5$ and $M=0.8$ for three values of $n$ with $G=0.01$; details of this calculation are given in Appendix \ref{app:numericsdecay}. There is excellent agreement between the numerically-derived prediction for $\sigma$ and the values obtained from the linear stability analysis.

Finally, we consider the case $n=0$. In $\eta<\eta_{b_0}$, the pressure perturbation is constant, $\mcp_1=c_0$, whilst in $\eta>\eta_{f_0}$, $\mcp_1=d_0$. In the interface region, the pressure perturbation is linear; $\mcp_1=a+b\eta$. The kinematic boundary conditions (\ref{eq:kinlintrail}, \ref{eq:kinlinlead}) furnish $\sigma=-1$. 
The thickness perturbation takes the form (see \ref{eq:h1p1rel})
\begin{equation}
\mch_1 = \frac{2 M b}{1-M} \eta^{-1}.
\end{equation}
The numerical results for axisymmetric perturbations to the self-similar solutions corroborated that $\sigma =-1$ (shown as a cross at $n=0$ in figure \ref{fig:dispersiondifference}). Axisymmetric perturbations decay faster than $\theta$-dependent perturbations with the error decaying proportional to $T^{-1}$ as has been found previously for unconfined gravity currents \citep{grundy1982eigenvalues,mathunjwa2006self}. For a different stability argument in the case of axisymmetric perturbations, see the Appendix of \citet{nordbotten2006similarity}.

\section{Vertically varying permeability} \label{sec:vertvarperm}
\begin{figure}
\centering
\includegraphics[width=0.95\columnwidth]{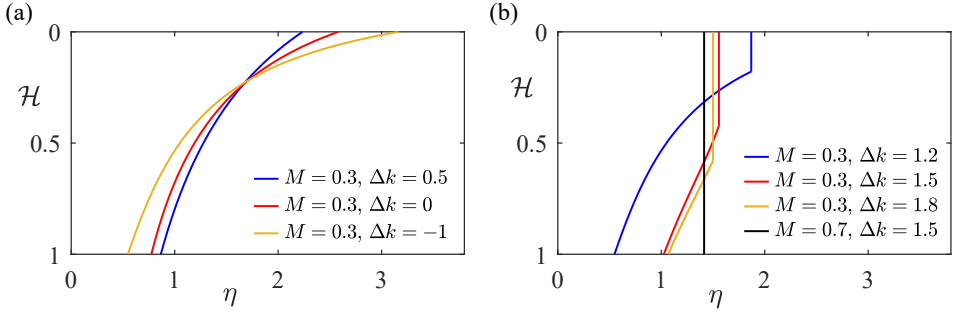}
\caption{Self-similar axisymmetric interface shapes in a layer with vertically varying permeability. (a) Small-$G$ analytic solutions \eqref{eq:G0intrusionvark} for different linear permeability variations \eqref{eq:linearperm}. (b) Small-$G$ analytic solutions in the case that there is a shock-like region.}
\label{fig:varpermG0shapes}
\end{figure}

The results obtained in previous sections generalise to layers in which the permeability varies vertically with $k=k(Z)$. The flow is axisymmetric and self-similar after buoyancy segregation. Here, we find the analytical solutions that arise in the small $G$ regime with a less viscous input fluid ($M<1$). One key difference to the uniform case is that shock-like regions in which the interface is relatively steep can occur even for $M<1$ (see \S \ref{subsec:shock}). Despite this, we show that at late times the axisymmetric solutions for any continuous permeability profile, $k(Z)$, are stable.

We define the dimensionless permeability as
\begin{equation}
\mathcal{K}(Z/H_0) = \frac{k(Z)}{\frac{1}{H_0} \int_0^{H_0} k(Z)\,\mathrm{d}Z},
\end{equation}
where the denominator is the mean permeability. We also define the dimensionless depth-integrated permeability \citep{hinton2018buoyancy},
\begin{equation}
\Phi(\mch) = \int_0^\mch \mathcal{K}(s) \, \mathrm{d} s.
\end{equation}
Note that $\Phi(0)=0$ and $\Phi(1)=1$. For a uniform layer, $\mathcal{K} \equiv 1$ and $\Phi(\mch)=\mch$.
The two mass conservation equations (\ref{eq:localmasscons_sim}, \ref{eq:fullmasscons_sim}) generalise to
\begin{equation} \label{eq:localmasscons_simvar}
\frac{\pd \mch}{\pd \tau}-\frac{\eta}{2} \frac{\pd \mch}{\pd \eta} - \tilde{\nabla} \cdot \big( \Phi(\mch) \tilde{\nabla} \mcp \big) = 0,
\end{equation}
\begin{equation} \label{eq:fullmasscons_simvar}
\tilde{\nabla} \cdot \Big[\big(M+(1-M)\Phi(\mch)\big) \tilde{\nabla} \mcp - G M (1-\Phi(\mch)) \tilde{\nabla} \mch   \Big] = 0.
\end{equation}
This model applies provided that the fluids have become segregated by buoyancy. In a uniform layer this applies at times given by \eqref{eq:segregatetime} with $k$ representing the vertical permeability. In a layer with vertically varying permeability and less viscous input fluid ($M<1$), the dimensional time for buoyancy segregation is adjusted to
\begin{equation} \label{eq:segtimevark}
T \gg \frac{\mu_a}{\Delta \rho g} \int_0^{H_0} \frac{1}{k_v(Z)} \mathrm{d} Z,
\end{equation}
where $k_v(Z)$ is the vertical permeability. In an isotropic, vertically heterogeneous layer, $k_v(Z)=k(Z)$. The time \eqref{eq:segtimevark} becomes singular if the permeability is zero at any height in the layer (and in this case our model does not apply). The time for the shallow approximation to apply is given by \eqref{eq:shallowtime}.

As for a uniform layer, there are self-similar axisymmetric solutions, $\mch_0=\mch_0(\eta)$, $\mcp=\mcp_0(\eta)$ to \eqref{eq:localmasscons_simvar}, \eqref{eq:fullmasscons_simvar}, with the pressure gradient at the top boundary given by
\begin{equation}
\frac{\mathrm{d} \mcp_0}{\mathrm{d} \eta} = \frac{-1}{\eta\big(M+(1-M)\Phi(\mch_0)\big)} + \frac{GM (1-\Phi(\mch_0))}{\big(M+(1-M)\Phi(\mch_0)\big)} \frac{\mathrm{d} \mch_0}{\mathrm{d} \eta},
\end{equation}
and $\mch_0$ satisfies the following ordinary differential equation
\begin{equation} \label{eq:mch0odevark}
-\frac{\eta}{2} \frac{\mathrm{d} \mch_0}{\mathrm{d} \eta} + \frac{1}{\eta} \frac{\mathrm{d}}{\mathrm{d} \eta} \Bigg(\frac{\Phi(\mch_0)}{M+(1-M) \Phi(\mch_0)} \Bigg) = \frac{1}{\eta} \frac{\mathrm{d}}{\mathrm{d} \eta} \Bigg(\eta \frac{G M \Phi(\mch_0)(1-\Phi(\mch_0))}{M+(1-M) \Phi(\mch_0)}\frac{\mathrm{d} \mch_0}{\mathrm{d} \eta} \Bigg).
\end{equation}
The interface gradient and pressure gradient at the leading contact line are
\begin{equation} \label{eq:gradleadvark}
\frac{\mathrm{d} \mch_0}{\pd \eta} = \frac{1}{\eta_{f_0} M G} - \frac{\eta_f}{2 G \mathcal{K}(0)}, \qquad \frac{\mathrm{d} \mcp_0}{\pd \eta} = \frac{-\eta_{f_0}}{2 \mathcal{K}(0)},
\end{equation}
respectively. Whilst at the trailing contact line
\begin{equation} \label{eq:gradtrailvark}
\frac{\mathrm{d} \mch_0}{\mathrm{d} \eta} = \frac{\eta_{b_0}}{2 M G \mathcal{K}(1)} - \frac{1}{\eta_{b_0} G}, \qquad \frac{\mathrm{d} \mcp_0}{\mathrm{d} \eta} = \frac{-1}{\eta_{b_0}}.
\end{equation}
The small-$G$ analytic solutions are given by neglecting the right-hand side of \eqref{eq:mch0odevark} from which we obtain the following implicit solution
\begin{align}
\mch_0&=1,  &   0<&\eta< \eta_{b_0},\\
\eta&=\frac{\sqrt{2M \mathcal{K}(\mch_0)}}{M+(1-M)\Phi(\mch_0)}, &  0<&\mch_0< 1, \label{eq:G0intrusionvark}  \\
\mch_0&=0, &  \eta_{f_0}<&\eta,
\end{align}
where the contact lines are
\begin{equation}
\eta_{f_0} = \sqrt{\frac{2  \mathcal{K}(0)}{M}}, \qquad \eta_{b_0} =\sqrt{2 M  \mathcal{K}(1)}.
\end{equation}
The associated pressure gradient in the interface region is given by
\begin{equation} \label{eq:dpdetavark1}
\frac{\mathrm{d} \mcp_0}{\mathrm{d} \eta} = -\frac{1}{\sqrt{2M \mathcal{K}(\mch_0)}}.
\end{equation}
In a uniform layer, the interface \eqref{eq:G0intrusionvark} is monotonic whenever $M<1$ but in a layer with a vertical permeability variation, turning points can arise even when the input fluid is less viscous than the ambient \citep{hinton2018buoyancy}. The solution \eqref{eq:G0intrusionvark} is only valid when the interface is monotonic. Otherwise, the interface has a turning point with buoyant input fluid lying below denser ambient fluid. This would invalidate the model, which assumed that the fluids have segregated owing to buoyancy.  For example, for a linear permeability structure with dimensionless variation, $\Delta k$,
\begin{equation} \label{eq:linearperm}
\mathcal{K}(s) = 1+ \Delta k (s-1/2),
\end{equation}
the interface \eqref{eq:G0intrusionvark} is monotonic if and only if \citep{hinton2018buoyancy}
\begin{equation}
M < \frac{(2-\Delta k)^2}{4 - 2 \Delta k + \Delta k^2}.
\end{equation}
Provided \eqref{eq:G0intrusionvark} is monotonic, the interface forms an axisymmetric self-similar intrusion, which extends along the upper boundary in a qualitatively identical fashion to the case of a uniform layer (see figure \ref{fig:varpermG0shapes}a). 
In the case that \eqref{eq:G0intrusionvark} has a turning point, a shock must be introduced; this regime is studied in \S \ref{subsec:shock}.

We apply our numerical method to layers with vertical variations in permeability for a wide range of initial conditions and find that the self-similar axisymmetric solutions are stable to both $\theta$-dependent and axisymmetric perturbations for any $G > 0$. Next, we generalise \S \ref{sec:stabilmech} to show how the pressure gradients suppress instabilities in a layer with vertically varying permeability (but no shock-like regions; for that case, see \S \ref{subsec:shock}). Between the contact lines, the magnitude of the pressure gradient in the ambient fluid is less than or equal to that in the input fluid owing to the contribution of the term $-G\pd \mch_0/\pd \eta$ (as in \S \ref{sec:stabilmech}). The interface gradient at the contact lines is discontinuous and the stability there is treated separately.

We can bound the location of the contact lines of the self-similar axisymmetric solution. In the case that the small-$G$ analytic solution \eqref{eq:G0intrusionvark} is monotonic, the contacts lines for $G > 0$ satisfy
\begin{equation} \label{eq:ineqconvark}
\eta_{f_0} \geq \sqrt{\frac{2  \mathcal{K}(0)}{M}}, \qquad \eta_{b_0} \leq \sqrt{2 M  \mathcal{K}(1)},
\end{equation}
with equality as $G\to 0^+$. These inequalities reflect the action of buoyancy to extend the interface. 

The gradient of the interface and the pressure at the upper boundary at the contact lines are given by \eqref{eq:gradleadvark} and \eqref{eq:gradtrailvark}. Thus, the pressure gradient in the ambient fluid just ahead of the leading contact line is given by
\begin{equation}
\frac{\mathrm{d}(\mcp_0-G \mch_0)}{\mathrm{d} \eta} = - \frac{1}{M \eta_{f_0}},
\end{equation}
whilst in the input fluid just upstream of the leading contact line, the pressure gradient is
\begin{equation}
\frac{\mathrm{d} \mcp_0}{\mathrm{d} \eta} = - \frac{\eta_{f_0}}{2 \mathcal{K}(0)},
\end{equation}
and the inequality \eqref{eq:ineqconvark}a ensures that the magnitude of the pressure gradient is larger in the input fluid, which stabilises the interface, with equality as $G\to 0^+$. 

The pressure gradient in the ambient fluid just ahead of the trailing contact line is given by
\begin{equation}
\frac{\mathrm{d}(\mcp_0-G \mch_0)}{\mathrm{d} \eta} = - \frac{\eta_{b_0}}{2 M \mathcal{K}(1)},
\end{equation}
whilst in the input fluid just upstream of the trailing contact line, the pressure gradient is
\begin{equation}
\frac{\mathrm{d} \mcp_0}{\mathrm{d} \eta} = - \frac{1}{\eta_{b_0}},
\end{equation}
and the inequality \eqref{eq:ineqconvark}b ensures that the magnitude of the pressure gradient is larger in the input fluid, which stabilises the interface.

\subsection{Shock-like regions} \label{subsec:shock}
In the case in which the small-$G$ interface \eqref{eq:G0intrusionvark} has a turning point, a shock must be introduced to vertically segregate the fluids. This shock may occupy part or all of the layer (see figure \ref{fig:varpermG0shapes}b).  By way of an example, we consider a linear variation in permeability with $\Delta k=1.5$ and viscosity ratio, $M=0.3$. The axisymmetric self-similar profile has a shock-like region near the top of the layer, even though $M<1$ (red line in figure \ref{fig:varpermG0shapes}b). The shock at the top of the layer is at location $\eta=\eta_s$ and is of thickness $\mch_s$, which satisfy
\begin{equation}
\eta_s=\sqrt{2F'(\mch_s)}, \qquad \eta_s^2 \mch_s = 2 F(\mch_s), \quad \text{where} \quad F(\mch) = \frac{\Phi(\mch)}{M+(1-M)\Phi(\mch)}.
\end{equation}
These arise from continuity with the solution \eqref{eq:G0intrusionvark}, which is valid in $\mch>\mch_s$, and mass conservation, respectively. The shock-like region may also extend across the entire layer (e.g. black line in figure \ref{fig:varpermG0shapes}b). For more complicated permeability profiles, there can be multiple shock-like regions separated by rarefaction-like regions \citep{hinton2018buoyancy}.

\begin{figure}
\centering
\includegraphics[width=0.95\columnwidth]{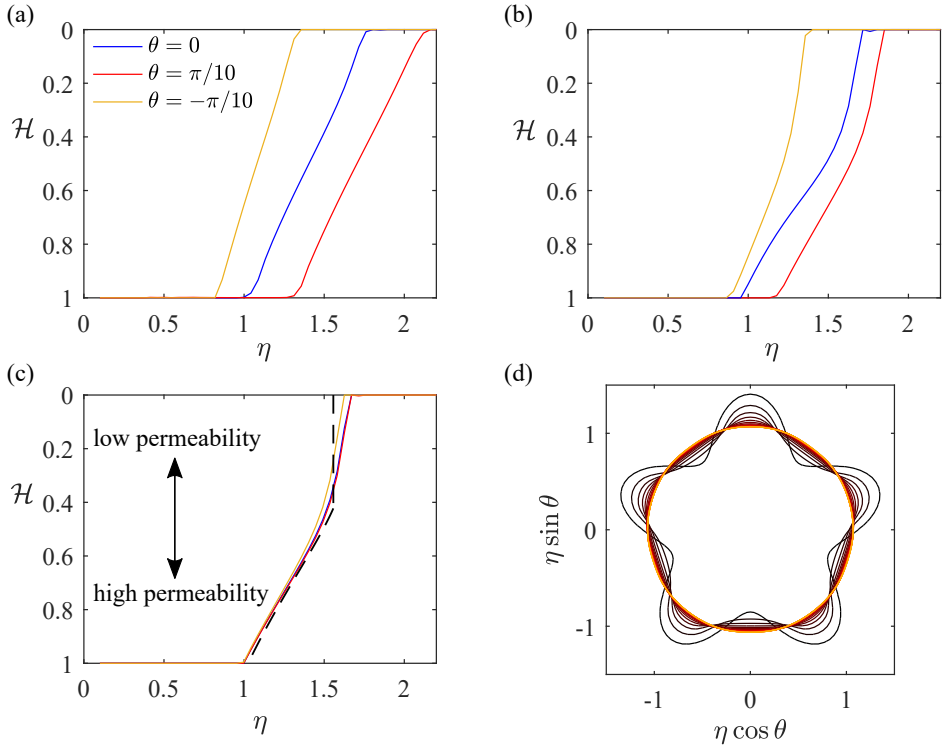}
\caption{Shutdown of mode five fingers in a layer with vertically varying permeability. We use a linear variation with $\Delta k=1.5$ and viscosity ratio, $M=0.3$, and $G=0.05$. Flow thickness at three radial cross-sections at (a) $\tau=0.5$, (b) $\tau=1.5$ and (c) $\tau=20$ (results calculated numerically; see \S \ref{sec:numerics}). In (c), the black dashed line shows the axisymmetric small-$G$ solution with a shock-like region in the upper portion of the layer. (d) Location of the trailing contact line $\eta=\eta_b(\theta,\tau)$ at unit time intervals between $\tau=0$ and $\tau=20$ (lighter colours correspond to later times).}
\label{fig:varperm}
\end{figure}

We apply our numerical method to layers with shock-like regions and find that the self-similar axisymmetric solutions are stable to both $\theta$-dependent and axisymmetric perturbations. An example is shown in figure \ref{fig:varperm} with a linear variation in permeability with $\Delta k=1.5$, viscosity ratio $M=0.3$, and $G=0.05$, demonstrating that mode five fingers are suppressed. The small-$G$ analytic solution is shown as a dashed line in in figure \ref{fig:varperm}c; there is a shock-like region in the top part of the layer. For $G>0$, the shock-like regions are smoothed by buoyancy so that they are not vertical \citep{hinton2018buoyancy}. The smoothed shock-like region has radial extent proportional to $G$ in the similarity coordinate $\eta$. Hence, in dimensional terms, the aspect ratio of the shock evolves with
\begin{equation}
\frac{L}{H_0} \sim G \Bigg(\frac{QT}{2 \pi \phi H_0^3} \Bigg)^{1/2},
\end{equation}
and eventually the shock-like region of the flow is long and thin (when the right-hand side becomes large) and the shallow flow approximation applies at sufficiently late times. This smoothing of the shock-like region also occurs when the shock encompasses the entire thickness of the layer (e.g. figure \ref{fig:varpermG0shapes}b). Thus the fluid-fluid interface is not cylindrical in the case. Instead, due to buoyancy, the interface becomes gradually shallower over time, which suppresses the classical instability.

The stability analysis generalises to the case in which there are shock-like regions.
At the top of the layer, the leading contact line for $G > 0$ satisfies
\begin{equation}
\eta_{f_0}\geq \eta_s \geq \sqrt{\frac{2 \mathcal{K}(0)}{M}}.
\end{equation}
The first inequality arises from buoyancy smoothing the interface beyond the shock location (see figure \ref{fig:varperm}c) whilst the second arises because the profile \eqref{eq:G0intrusionvark} has a turning point in the shock-like region and the shock conserves mass so it must be ahead of this profile at $Z=0$. Hence \eqref{eq:ineqconvark}a applies even when there is a shock-like region at the top of the layer. A similar argument applies when there is a shock-like region at the bottom of the layer and \eqref{eq:ineqconvark}b remains correct. 

The gradient of the interface and the pressure at the upper boundary at the contact lines are given by \eqref{eq:gradleadvark} and \eqref{eq:gradtrailvark}, even when there are shock-like regions. Hence the stability argument of \S \ref{sec:vertvarperm} applies.

In summary, the self-similar axisymmetric solutions in a layer with vertically varying permeability are stable owing to the same mechanism as in a uniform layer. The less viscous input fluid intrudes into the ambient fluid sufficiently far that the destabilising increase in the magnitude of the pressure gradient associated with the viscosity contrast is smoothed over the extent of the interface. Buoyancy segregation corresponds to a monotonic interface and the input fluid may be forced into the low permeability region \citep[see also][]{debbabi2018impact}. We note that there may be significant intermingling of the fluids in the transient evolution to the buoyancy-segregated flow \citep[e.g.][]{huppert2013competition}. There may also be a competition between viscous fingering and heterogeneity-driven fingering prior to buoyancy segregation \citep{de1997viscous}.

\section{Discussion and conclusion} \label{sec:conc}
We have examined the stability of the horizontal flow of one fluid injected into another fluid of greater viscosity and density. When there is a sharp interface and the two fluids have segregated owing to buoyancy, the flow evolves in an axisymmetric self-similar fashion. Whilst Saffman-Taylor fingers may occur prior to buoyancy segregation, we have demonstrated that the buoyancy-segregated self-similar flow is stable to both axisymmetric and angular-dependent perturbations. The input fluid forms an intrusion with large radial extent neighbouring the upper boundary. This disperses the destabilising pressure gradient jump associated with the viscosity contrast between the fluids that drives the classical instability. The mechanism generalises to layers with a vertical variation in permeability and anisotropic layers with different horizontal and vertical permeability. After buoyancy segregation, the flow in a heterogeneous layer is always stable even though the interface may contain steep shock-like regions owing to slow flow in the low permeability zone. The time for buoyancy segregation can be long if the porous layers has zones of very low permeability.

In future work, these ideas could be extended to viscous displacements in horizontal channels as is relevant to mantle plumes \citep{schoonman2017radial}. The base self-similar flow was found for a line source and a point source by \citet{zheng2015viscous} and \citet{hinton2020axisymmetric} but the stability has not yet been investigated. The analysis would be more complicated than the present work owing to the effect of the no-slip boundaries and the shear flow \citep{snyder1998flow,john2013variable}. 

In the context of CO\textsubscript{2} sequestration, viscous fingering is undesirable as it may reduce storage efficiency. Our study could be usefully extended to consider strategies to vary the injection rate to avoid viscous fingering. Given that the interface is stable after buoyancy segregation (even for relatively high injection rates) one could analyse how to vary the input flux during the pre-segregation transient to ensure that fingers never occur.

\section*{Acknowledgements}
E.M.H. is grateful to the University of Melbourne for the award of a Harcourt-Doig research fellowship.

\section*{Declaration of interests}
The authors report no conflict of interest.

\appendix

\section{Numerical method} \label{app:numerics}
In this appendix, the numerical method for calculating the evolution of the fluid-fluid interface and the pressure at the upper boundary is described. The two dependent variables, $\mch(\eta, \theta, \tau)$, and $\mcp(\eta, \theta, \tau)$ are calculated from equations \eqref{eq:localmasscons_sim} and \eqref{eq:fullmasscons_sim} with boundary conditions \eqref{eq:bc1_sim} and \eqref{eq:bc2_sim} using a finite difference method. 
\par
The numerical scheme requires initial data for the interface shape at the reference time ($\tau=0$). The initial condition was chosen to satisfy the mass conservation condition \eqref{eq:transformglmasscon}. The pressure at the upper boundary, $\mcp$, and interface shape, $\mch$, must satisfy \eqref{eq:fullmasscons_sim} initially since the fluid is incompressible and the upper and lower boundaries are impermeable. They must also satisfy the boundary conditions \eqref{eq:bc1_sim} and \eqref{eq:bc2_sim}. Admissible choices for the initial conditions are constructed by first selecting any $\mch$, which satisfies the boundary conditions and then solving \eqref{eq:fullmasscons_sim} (described below) to obtain an initial form for $\mcp$. 
\par
The independent variables are discretised on an annular domain with $(\eta,\theta,\tau) \in [\eta_l,\eta_r] \times [0, 2 \pi) \times [0,\tau_1]$, where $\tau_1$ is the time to which the simulation is run, and the inner and outer radii are chosen such that the domain fully incorporates the contact lines and $\eta_l$ is taken to be small but non-zero so that the boundary condition \eqref{eq:bc1_sim} may be applied. Similarly $\eta_r$ is chosen to be large and \eqref{eq:bc2_sim} is applied. Typically we used $\eta_l=0.01$ and $\eta_r=10$ and we confirmed that changes to these values led to imperceptible differences to the results obtained. The discretised variables are
\begin{align}
\eta_i &= \eta_l + i \Delta \eta, &   i&=0 \ldots L, & \Delta \eta &= \frac{\eta_r-\eta_l}{L}, \\
\theta_j &= j \Delta \theta, &   j&=0 \ldots M-1, & \Delta \theta &= \frac{2 \pi}{M}, \\
\tau_k &= k \Delta \tau, &   k&=0 \ldots N, & \Delta \tau &= \frac{\tau_1}{N},
\end{align}
and we write $\mch_{i,j,k}$ and $\mcp_{i,j,k}$ to denote the approximations of the dependent variables.
\par
The numerical procedure is as follows. First, the given initial condition for $\mch$ and $\mcp$ is discretized. At each subsequent time step, the interface shape is updated via an adapted midpoint (second-order Runge-Kutta) method. The interface height at the midpoint between timesteps is given by (using \ref{eq:localmasscons_sim})
\begin{equation}
\mch_{i,j,k+1/2} = \mch_{i,j,k} + \frac{\Delta \tau}{2} \Bigg[ \frac{\eta}{2} \frac{\partial \mch}{\partial \eta} + \nabla \cdot (\mch \nabla \mcp) \Bigg]_{i,j,k},
\end{equation}
where the term in square brackets is calculated using central differences in the two spatial coordinates. The pressure at the midpoint, $\mcp_{i,j,k+1/2}$ is obtained from \eqref{eq:fullmasscons_sim} using $\mch=\mch_{i,j,k+1/2}$ and applying a five-point difference formula and the relaxation method to solve the steady problem. The timestepping is completed with
\begin{equation}
\mch_{i,j,k+1} = \mch_{i,j,k} + \Delta \tau \Bigg[ \frac{\eta}{2} \frac{\partial \mch}{\partial \eta} + \nabla \cdot (\mch \nabla \mcp) \Bigg]_{i,j,k+1/2},
\end{equation}
and $\mcp_{i,j,k+1}$ is then obtained in an identical fashion to $\mcp_{i,j,k+1/2}$. We use $\Delta \tau=0.1 \Delta \eta^2$.
\par
Various checks were used to verify the accuracy of the numerical method. First, we confirmed that mass conservation \eqref{eq:transformglmasscon} was satisfied to within $0.05$ \% at all times over a large range of the parameters; $0<M<10$ and $0 < G <10$. Second, we used the axisymmetric similarity solution as the initial condition and found that the solution was steady. Finally, we confirmed that the results always converged to the axisymmetric similarity solution.

\subsection{Estimating the decay rate ($\sigma$)} \label{app:numericsdecay}
In this section we describe how the decay rate, $\sigma$, of linear perturbations to the axisymmetric self-similar flow can be estimated using the numerical method. We use the following initial condition for the numerical method
\begin{equation}
\mathcal{H}(\eta,\theta,\tau=0) = \mathcal{H}_0\bigg( \frac{\eta}{1+0.1\cos(n\theta)} \bigg),
\end{equation}
for mode $n$ perturbations, where $\mathcal{H}_0(\eta)$ is the axisymmetric self-similar solution. We integrate forward in time as described in Appendix \ref{app:numerics}. The error between the resultant numerical solution, $\mathcal{H}(\eta,\theta,\tau)$ and the axisymmetric solution, $\mathcal{H}_0(\eta)$ is calculated at each time via
\begin{equation}
E(\tau) = \int_0^{\eta_r} \int_0^{2\pi} \sqrt{\mathcal{H}^2-\mathcal{H}_0^2} \, \mathrm{d} \theta \mathrm{d} \eta.
\end{equation}
At sufficiently late times, the linear stability analysis of \S \ref{sec:linstabil} applies. To predict the decay rate, $\sigma$, we best-fit a straight line to $\log E$ as a function of $\tau \in [5,10]$. Some of the results for $G=0.01$ are shown as crosses in figure \ref{fig:decayrates}.

\bibliographystyle{jfm}
\bibliography{jfm2}

\begin{thebibliography}{38}
\expandafter\ifx\csname natexlab\endcsname\relax\def\natexlab#1{#1}\fi
\def\au#1{#1} \def\ed#1{#1} \def\yr#1{#1}\def\at#1{#1}\def\jt#1{\textit{#1}}
  \def\bt#1{#1}\def\bvol#1{\textbf{#1}} \def\vol#1{#1} \def\pg#1{#1}
  \def\publ#1{#1}\def\arxiv#1{#1}\def\org#1{#1}\def\st#1{\textit{#1}}

\bibitem[Al-Housseiny {\em et~al.\/}(2012)Al-Housseiny, Tsai \&
  Stone]{al2012control}
{\sc \au{Al-Housseiny, T.T.}, \au{Tsai, P.A.} \& \au{Stone, H.A.}} \yr{2012}
  \at{Control of interfacial instabilities using flow geometry}.  \jt{Nat.
  Phys.}  \bvol{8}~(10),  \pg{747--750}.

\bibitem[Bachu(2015)]{bachu2015review}
{\sc \au{Bachu, S.}} \yr{2015}  \at{{Review of CO2 storage efficiency in deep
  saline aquifers}}.  \jt{Int. J. Greenh. Gas Control.}  \bvol{40},
  \pg{188--202}.

\bibitem[Benham {\em et~al.\/}(2022)Benham, Neufeld \&
  Woods]{benham2022axisymmetric}
{\sc \au{Benham, G.P.}, \au{Neufeld, J.A.} \& \au{Woods, A.W.}} \yr{2022}
  \at{Axisymmetric gravity currents in anisotropic porous media}.  \jt{arXiv
  preprint arXiv:2202.01165} .

\bibitem[Bernoff \& Witelski(2002)]{bernoff2002linear}
{\sc \au{Bernoff, A.J.} \& \au{Witelski, T.P.}} \yr{2002}  \at{Linear stability
  of source-type similarity solutions of the thin film equation}.  \jt{Appl.
  Math. Lett.}  \bvol{15}~(5),  \pg{599--606}.

\bibitem[Chertock(2002)]{chertock2002stability}
{\sc \au{Chertock, A.}} \yr{2002}  \at{On the stability of a class of
  self-similar solutions to the filtration-absorption equation}.  \jt{Eur. J.
  Appl. Math.}  \bvol{13}~(2),  \pg{179--194}.

\bibitem[De~Wit \& Homsy(1997)]{de1997viscous}
{\sc \au{De~Wit, A.} \& \au{Homsy, G.M.}} \yr{1997}  \at{{Viscous fingering in
  periodically heterogeneous porous media. II. Numerical simulations}}.  \jt{J.
  Chem. Phys.}  \bvol{107}~(22),  \pg{9619--9628}.

\bibitem[Debbabi {\em et~al.\/}(2018)Debbabi, Jackson, Hampson \&
  Salinas]{debbabi2018impact}
{\sc \au{Debbabi, Y.}, \au{Jackson, M.D.}, \au{Hampson, G.J.} \& \au{Salinas,
  P.}} \yr{2018}  \at{Impact of the buoyancy--viscous force balance on
  two-phase flow in layered porous media}.  \jt{Transp. Porous Media}
  \bvol{124}~(1),  \pg{263--287}.

\bibitem[Grundy \& McLaughlin(1982)]{grundy1982eigenvalues}
{\sc \au{Grundy, R.E.} \& \au{McLaughlin, R.}} \yr{1982}  \at{{Eigenvalues of
  the Barenblatt-Pattle similarity solution in nonlinear diffusion}}.
  \jt{Proc. R. Soc. A: Math. Phys. Eng. Sci.}  \bvol{383}~(1784),
  \pg{89--100}.

\bibitem[Guo {\em et~al.\/}(2016)Guo, Zheng, Celia \&
  Stone]{guo2016axisymmetric}
{\sc \au{Guo, B.}, \au{Zheng, Z.}, \au{Celia, M.A.} \& \au{Stone, H.A.}}
  \yr{2016}  \at{Axisymmetric flows from fluid injection into a confined porous
  medium}.  \jt{Phys. Fluids}  \bvol{28}~(2),  \pg{022107}.

\bibitem[Hesse {\em et~al.\/}(2007)Hesse, Tchelepi, Cantwel \&
  Orr]{hesse2007gravity}
{\sc \au{Hesse, M.A.}, \au{Tchelepi, H.A.}, \au{Cantwel, B.J.} \& \au{Orr,
  F.M.}} \yr{2007}  \at{Gravity currents in horizontal porous layers:
  transition from early to late self-similarity}.  \jt{J. Fluid Mech.}
  \bvol{577},  \pg{363--383}.

\bibitem[Hinton(2020)]{hinton2020axisymmetric}
{\sc \au{Hinton, E.M.}} \yr{2020}  \at{Axisymmetric viscous flow between two
  horizontal plates}.  \jt{Phys. Fluids}  \bvol{32}~(6),  \pg{063104}.

\bibitem[Hinton \& Woods(2018)]{hinton2018buoyancy}
{\sc \au{Hinton, E.M.} \& \au{Woods, A.W.}} \yr{2018}  \at{Buoyancy-driven flow
  in a confined aquifer with a vertical gradient of permeability}.  \jt{J.
  Fluid Mech.}  \bvol{848},  \pg{411--429}.

\bibitem[Homsy(1987)]{homsy1987viscous}
{\sc \au{Homsy, G.M.}} \yr{1987}  \at{Viscous fingering in porous media}.
  \jt{Annu. Rev. Fluid Mech.}  \bvol{19}~(1),  \pg{271--311}.

\bibitem[Huppert {\em et~al.\/}(2013)Huppert, Neufeld \&
  Strandkvist]{huppert2013competition}
{\sc \au{Huppert, H.E.}, \au{Neufeld, J.A.} \& \au{Strandkvist, C.}} \yr{2013}
  \at{The competition between gravity and flow focusing in two-layered porous
  media}.  \jt{J. Fluid Mech.}  \bvol{720},  \pg{5--14}.

\bibitem[Huppert \& Pegler(2022)]{huppert2022fate}
{\sc \au{Huppert, H.E.} \& \au{Pegler, S.S.}} \yr{2022}  \at{The fate of
  continuous input of relatively heavy fluid at the base of a porous medium}.
  \jt{J. Fluid Mech.}  \bvol{932}.

\bibitem[Huppert \& Woods(1995)]{huppert1995gravity}
{\sc \au{Huppert, H.E.} \& \au{Woods, A.W.}} \yr{1995}  \at{Gravity-driven
  flows in porous layers}.  \jt{J. Fluid Mech.}  \bvol{292},  \pg{55--69}.

\bibitem[John {\em et~al.\/}(2013)John, Oliveira, Heussler \&
  Meiburg]{john2013variable}
{\sc \au{John, M.O.}, \au{Oliveira, R.M.}, \au{Heussler, F.H.C.} \&
  \au{Meiburg, E.}} \yr{2013}  \at{{Variable density and viscosity, miscible
  displacements in horizontal Hele-Shaw cells. Part 2. Nonlinear simulations}}.
   \jt{J. Fluid Mech.}  \bvol{721},  \pg{295--323}.

\bibitem[Juanes {\em et~al.\/}(2010)Juanes, MacMinn \&
  Szulczewski]{juanes2010footprint}
{\sc \au{Juanes, R.}, \au{MacMinn, C.W.} \& \au{Szulczewski, M.L.}} \yr{2010}
  \at{{The footprint of the CO2 plume during carbon dioxide storage in saline
  aquifers: storage efficiency for capillary trapping at the basin scale}}.
  \jt{Transp. Porous Media}  \bvol{82}~(1),  \pg{19--30}.

\bibitem[Kowal(2021)]{kowal2021viscous}
{\sc \au{Kowal, K.N.}} \yr{2021}  \at{Viscous banding instabilities: non-porous
  viscous fingering}.  \jt{J. Fluid Mech.}  \bvol{926},  \pg{A4}.

\bibitem[Kowal \& Worster(2019)]{kowal2019stability}
{\sc \au{Kowal, K.N.} \& \au{Worster, M.G.}} \yr{2019}  \at{{Stability of
  lubricated viscous gravity currents. Part 2. Global analysis and
  stabilisation by buoyancy forces}}.  \jt{J. Fluid Mech.}  \bvol{871},
  \pg{1007--1027}.

\bibitem[Lyle {\em et~al.\/}(2005)Lyle, Huppert, Hallworth, Bickle \&
  Chadwick]{lyle2005axisymmetric}
{\sc \au{Lyle, S.}, \au{Huppert, H.E.}, \au{Hallworth, M.}, \au{Bickle, M.} \&
  \au{Chadwick, A.}} \yr{2005}  \at{Axisymmetric gravity currents in a porous
  medium}.  \jt{J. Fluid Mech.}  \bvol{543},  \pg{293--302}.

\bibitem[Mathunjwa \& Hogg(2006)]{mathunjwa2006self}
{\sc \au{Mathunjwa, J.S.} \& \au{Hogg, A.J.}} \yr{2006}  \at{{Self-similar
  gravity currents in porous media: linear stability of the Barenblatt--Pattle
  solution revisited}}.  \jt{Eur. J. Mech. B Fluids}  \bvol{25}~(3),
  \pg{360--378}.

\bibitem[Newman(1984)]{newman1984lyapunov}
{\sc \au{Newman, W.I.}} \yr{1984}  \at{{A Lyapunov functional for the evolution
  of solutions to the porous medium equation to self-similarity. I}}.  \jt{J.
  Math. Phys.}  \bvol{25}~(10),  \pg{3120--3123}.

\bibitem[Nijjer {\em et~al.\/}(2018)Nijjer, Hewitt \&
  Neufeld]{nijjer2018dynamics}
{\sc \au{Nijjer, J.S.}, \au{Hewitt, D.R.} \& \au{Neufeld, J.A.}} \yr{2018}
  \at{The dynamics of miscible viscous fingering from onset to shutdown}.
  \jt{J. Fluid Mech.}  \bvol{837},  \pg{520--545}.

\bibitem[Nijjer {\em et~al.\/}(2022)Nijjer, Hewitt \&
  Neufeld]{nijjer2022horizontal}
{\sc \au{Nijjer, J.S.}, \au{Hewitt, D.R.} \& \au{Neufeld, J.A.}} \yr{2022}
  \at{Horizontal miscible displacements through porous media: the interplay
  between viscous fingering and gravity segregation}.  \jt{J. Fluid Mech.}
  \bvol{935},  \pg{A14}.

\bibitem[Nordbotten \& Celia(2006)]{nordbotten2006similarity}
{\sc \au{Nordbotten, J.M.} \& \au{Celia, M.A.}} \yr{2006}  \at{Similarity
  solutions for fluid injection into confined aquifers}.  \jt{J. Fluid Mech.}
  \bvol{561},  \pg{307--327}.

\bibitem[Paterson(1981)]{paterson1981radial}
{\sc \au{Paterson, L.}} \yr{1981}  \at{{Radial fingering in a Hele Shaw cell}}.
   \jt{J. Fluid Mech.}  \bvol{113},  \pg{513--529}.

\bibitem[Pattle(1959)]{pattle1959diffusion}
{\sc \au{Pattle, R.E.}} \yr{1959}  \at{Diffusion from an instantaneous point
  source with a concentration-dependent coefficient}.  \jt{Q. J. Mech. Appl.
  Math.}  \bvol{12}~(4),  \pg{407--409}.

\bibitem[Pegler {\em et~al.\/}(2014)Pegler, Huppert \&
  Neufeld]{pegler2014fluid}
{\sc \au{Pegler, S.S.}, \au{Huppert, H.E.} \& \au{Neufeld, J.A.}} \yr{2014}
  \at{Fluid injection into a confined porous layer}.  \jt{J. Fluid Mech.}
  \bvol{745},  \pg{592--620}.

\bibitem[Riaz \& Meiburg(2003)]{riaz2003three}
{\sc \au{Riaz, A.} \& \au{Meiburg, E.}} \yr{2003}  \at{Three-dimensional
  miscible displacement simulations in homogeneous porous media with gravity
  override}.  \jt{J. Fluid Mech.}  \bvol{494},  \pg{95--117}.

\bibitem[Saffman \& Taylor(1958)]{saffman1958penetration}
{\sc \au{Saffman, P.G.} \& \au{Taylor, G.I.}} \yr{1958}  \at{{The penetration
  of a fluid into a porous medium or Hele-Shaw cell containing a more viscous
  liquid}}.  \jt{Proc. R. Soc. A: Math. Phys. Eng. Sci.}  \bvol{245}~(1242),
  \pg{312--329}.

\bibitem[Schoonman {\em et~al.\/}(2017)Schoonman, White \&
  Pritchard]{schoonman2017radial}
{\sc \au{Schoonman, C.M.}, \au{White, N.J.} \& \au{Pritchard, D.}} \yr{2017}
  \at{{Radial viscous fingering of hot asthenosphere within the Icelandic plume
  beneath the North Atlantic Ocean}}.  \jt{Earth Planet. Sci. Lett.}
  \bvol{468},  \pg{51--61}.

\bibitem[Sharma {\em et~al.\/}(2020)Sharma, Nand, Pramanik, Chen \&
  Mishra]{sharma2020control}
{\sc \au{Sharma, V.}, \au{Nand, S.}, \au{Pramanik, S.}, \au{Chen, C.-Y.} \&
  \au{Mishra, M.}} \yr{2020}  \at{Control of radial miscible viscous
  fingering}.  \jt{J. Fluid Mech.}  \bvol{884}.

\bibitem[Snyder \& Tait(1998)]{snyder1998flow}
{\sc \au{Snyder, D.} \& \au{Tait, S.}} \yr{1998}  \at{A flow-front instability
  in viscous gravity currents}.  \jt{J. Fluid Mech.}  \bvol{369},  \pg{1--21}.

\bibitem[Tchelepi(1994)]{tchelepi1994viscous}
{\sc \au{Tchelepi, H.A.}} \yr{1994}  \at{Viscous fingering, gravity segregation
  and permeability heterogeneity in two-dimensional and three-dimensional
  flows}. PhD thesis, Stanford University.

\bibitem[Zheng {\em et~al.\/}(2015{\natexlab{{\em a\/}}})Zheng, Kim \&
  Stone]{zheng2015controlling}
{\sc \au{Zheng, Z.}, \au{Kim, H.} \& \au{Stone, H.A.}} \yr{2015{\natexlab{{\em
  a\/}}}}  \at{Controlling viscous fingering using time-dependent strategies}.
  \jt{Phys. Rev. Lett.}  \bvol{115}~(17),  \pg{174501}.

\bibitem[Zheng {\em et~al.\/}(2015{\natexlab{{\em b\/}}})Zheng, Rongy \&
  Stone]{zheng2015viscous}
{\sc \au{Zheng, Z.}, \au{Rongy, L.} \& \au{Stone, H.A.}}
  \yr{2015{\natexlab{{\em b\/}}}}  \at{Viscous fluid injection into a confined
  channel}.  \jt{Phys. Fluids}  \bvol{27}~(6),  \pg{062105}.

\bibitem[Zheng \& Stone(2022)]{zheng2022influence}
{\sc \au{Zheng, Z.} \& \au{Stone, H.A.}} \yr{2022}  \at{{The Influence of
  Boundaries on Gravity Currents and Thin Films: Drainage, Confinement,
  Convergence, and Deformation Effects}}.  \jt{Annu. Rev. Fluid Mech.}
  \bvol{54},  \pg{27--56}.

\end{thebibliography}

\end{document}